\documentclass[a4paper,fleqn,usenatbib]{mnras}

\usepackage[T1]{fontenc}
\usepackage{times}
\usepackage{mathptmx}

\usepackage{graphicx}	
\usepackage{amsmath}	
\usepackage{amssymb}	
\usepackage{ulem}  

\newcommand{\Msun}{M$_\text{\sun}$ }

\title[Supermassive black holes in IllustrisTNG]{Supermassive black holes and their feedback effects in the IllustrisTNG simulation}

\author[R. Weinberger et al.]{%
Rainer Weinberger$^{1}$\thanks{E-mail: rainer.weinberger@h-its.org},
Volker Springel$^{1,2,3}$,
R\"udiger Pakmor$^{1}$,
Dylan Nelson$^{3}$, \newauthor
Shy Genel$^{4,5}$,
Annalisa Pillepich$^{6}$,
Mark Vogelsberger$^{7}$,
Federico Marinacci$^{7}$, \newauthor
Jill Naiman$^{8}$,
Paul Torrey$^{7}$,
Lars Hernquist$^{8}$
\\%
$^{1}$Heidelberg Institute for Theoretical Studies, Schloss-Wolfsbrunnenweg 35, D-69118 Heidelberg, Germany\\
$^{2}$Zentrum f\"ur Astronomie der Universit\"at Heidelberg, ARI, M\"onchhofstrasse 12-14, D-69120 Heidelberg, Germany\\
$^{3}$Max-Planck-Institut f\"ur Astrophysik, Karl-Schwarzschild-Str. 1, D-85748, Garching, Germany\\
$^{4}$Center for Computational Astrophysics, Flatiron Institute, 162 Fifth Avenue, New York, NY 10010, USA\\
$^{5}$Columbia Astrophysics Laboratory, Columbia University, 550 West 120th Street, New York, NY 10027, USA\\
$^{6}$Max-Planck-Institut f\"ur Astronomie, K\"onigstuhl 17, D-69117 Heidelberg, Germany\\
$^{7}$Department of Physics, Kavli Institute for Astrophysics and Space Research, MIT, Cambridge, MA 02139, USA\\
$^{8}$Harvard-Smithsonian Center for Astrophysics, 60 Garden Street, Cambridge, MA 02138, USA
}

\date{Accepted XXX. Received YYY; in original form ZZZ}

\pubyear{2018}

\begin{document}
\label{firstpage}
\pagerange{\pageref{firstpage}--\pageref{lastpage}}
\maketitle

\begin{abstract} 
  We study the population of supermassive black holes (SMBHs) and their effects on massive central galaxies in the IllustrisTNG cosmological hydrodynamical simulations of galaxy formation. The employed model for SMBH growth and feedback assumes a two-mode scenario in which the feedback from active galactic nuclei occurs through a kinetic, comparatively efficient mode at low accretion rates relative to the Eddington limit, and in the form of a thermal, less efficient mode at high accretion rates.  We show that the quenching of massive central galaxies happens coincidently with kinetic-mode feedback, consistent with the notion that active supermassive black cause the low specific star formation rates observed in massive galaxies. However, major galaxy mergers are not responsible for initiating most of the quenching events in our model. Up to black hole masses of about $10^{8.5}\,{\rm M}_\text{\sun}$, the dominant growth channel for SMBHs is in the thermal mode. Higher mass black holes stay mainly in the kinetic mode and gas accretion is self-regulated via their feedback, which causes their Eddington ratios to drop, with SMBH mergers becoming the main channel for residual mass growth. As a consequence, the quasar luminosity function is dominated by rapidly accreting, moderately massive black holes in the thermal mode. We show that the associated growth history of SMBHs produces a low-redshift quasar luminosity function and a redshift zero black hole mass -- stellar bulge mass relation in good agreement with observations, whereas the simulation tends to over-predict the high-redshift quasar luminosity function.  \end{abstract}




\section{Introduction}

It is now well established that most if not all massive galaxies host supermassive black holes (SMBHs). If the growth of SMBHs is dominated by gas accretion, the corresponding energy released per unit volume is quite substantial \citep{Soltan1982} and matches the integrated emission from the quasar luminosity function, supporting that this is the primary growth channel. If only a small fraction of the released energy couples to the gas of the host galaxy, the impact of SMBHs on their host galaxies can be significant \citep{King2003, DiMatteo+2005}. The relatively tight scaling relations between supermassive black hole masses and properties of their host galaxies \citep[e.g.][]{Ferrarese2000} furthermore point towards a mutual influence on each other, thereby establishing some form of co-evolution.

Recent simulations that model the formation of massive galaxies rely on feedback effects from SMBHs to reproduce the properties of massive galaxies \citep{Springel+2005, Bower+2006, Croton+2006, Dubois+2013, Martizzi+2014, Choi+2015, Somerville+Dave2015}. However, even with presently available computational resources, it is not possible to model these effects from first principles. Instead, sub-resolution models are applied which typically measure gas properties at resolved scales, translate them via simplified analytic models to a black hole accretion rate, and inject feedback energy with some assumed efficiency into the surrounding gas on resolved scales. While these models contain tuneable parameters which are usually set such that the simulations reproduce the stellar properties of simulated galaxies and the relation between black hole mass and stellar mass, the calculations are able to {\it additionally} reproduce a variety of other (unconstrained) observable properties of active galactic nuclei (AGNs) \citep[e.g.][]{Sijacki+2015, Volonteri+2016}.  In this way, such models allow detailed insights into how the different growth and feedback processes of galaxies and their SMBHs are intertwined.

Observationally, a number of important properties of the SMBH population can be inferred. First, the luminosity function, which can be measured up to high redshift, gives insights about the mass growth via gas accretion over cosmic time \citep{Hopkins+2007, Shankar+2009, Ueda+2014, Lacy+2015}.  Using estimates for black hole masses and accretion rates, it is also possible to infer the distribution of Eddington ratios \citep{Schulze+2015, Georgakakis+2017}, or alternatively, the specific accretion rate distribution \citep{Aird+2017b, Aird+2017}, which constrains the state of the accretion disc over cosmic time \citep{Weigel+2017}. Furthermore, relating the SMBH properties to the galaxy properties, e.g.~the SMBH mass with the bulge mass (or velocity dispersion) of the host \citep{Magorrian+1998, Tremaine+2002, Haerig+Rix2004, McConnell+Ma2013, Kormendy+Ho2013, Graham+Scott2015, Reines+Volonteri2015, Savorgnan+2016}, indicates a connection between the two objects. Yet it remains debated whether this is an indication for the feedback regulated nature of SMBH growth \citep{King2003}, or just a manifestation of a common assembly history \citep{Peng2007, Hirschmann+2010, Jahnke+Maccio2011, Angles-Alcazar+2013}.

The comparatively accurate description of hierarchical structure formation obtained by cosmological simulations can be used to turn the statistical properties of the predicted SMBH population into powerful tests of the SMBH model adopted in a simulation \citep{DiMatteo+2008, DeGraf+2012, Hirschmann+2014, Sijacki+2015, Volonteri+2016}, even though the observations are subject to large uncertainties and selection biases.  In addition, it is possible to exploit the rich information contained in simulations to test specific scenarios for the SMBH -- galaxy coevolution, such as the role of major mergers \citep{Springel+2005, DiMatteo+2005, Hopkins+2006, Sparre+Springel2016b, Sparre+Springel2017, Pontzen+2017} or the interplay with other effects, like stellar feedback in galaxies of specific masses \citep{Dubois+2015, Habouzit+2016, Habouzit+2017, Bower+2017}.

In high-resolution simulations of representative cosmological volumes \citep{Vogelsberger+2014, Schaye+2015, Dubois+2016, Khandai+2015, Angles-Alcazar+2017, Tremmel+2017} it is possible to track the evolution of SMBHs and their host galaxies through cosmic time and relate the host galaxy transformations, for example quenching  or changes in morphology, to the SMBH properties.   Using a large number of simulated systems, it is possible to obtain statistical information about the diversity of the evolutionary paths of galaxies, which can reveal new insights about the SMBH -- galaxy coevolution.

In this paper, we use ``The Next Generation Illustris'' (IllustrisTNG) simulations to study the co-evolution of SMBHs and galaxies. In the introductory papers of the project \citep{Marinacci+2017, Naiman+2017, Nelson+2017, Pillepich+2017b, Springel+2017}, we have shown that IllustrisTNG reproduces a diverse range of observables remarkably well, in particular massive galaxies have significantly improved stellar \citep{Nelson+2017,Pillepich+2017b, Genel+2017} and gas properties \citep{Vogelsberger+2017, Marinacci+2017, Weinberger+2017} compared to the predecessor Illustris simulation \citep{Vogelsberger+2014,Genel+2014}. Based on this encouraging progress, we study the origin of the quenched, massive central galaxies and the role of SMBHs for their evolution. We focus in particular on the energetics of the AGN feedback and investigate the role of (major) mergers for quenching, black hole mass growth, and AGN activity.

In Section~\ref{sec:IllustrisTNG}, we present the IllustrisTNG simulations and briefly describe the numerical methods and astrophysical models used. The galaxy population of IllustrisTNG in terms of star formation rate and energetics of different feedback modes is presented in Section~\ref{sec:GalaxyPopulation}, and linked to the SMBH population in Section~\ref{sec:SMBH}. We discuss the results in Section~\ref{sec:discussion}, and conclude in Section~\ref{sec:Conclusion}.

\section{The Illustris TNG Simulations}
\label{sec:IllustrisTNG}

\begin{table}
\begin{center}
\begin{tabular}{c l c c c}
\hline
 Simulation name &  & TNG100 & TNG300 & TNG300-2\\
\hline
$N_{\rm cells}$ &  & $1820^3$ & $2500^3$ & $1250^3$ \\
$L_{\rm box}$ & [Mpc] & $111$ & $303$ & $303$ \\
$m_{\rm target, gas}$ & [$10^6$~\Msun] & $1.4$& $11$ & $88$ \\
$m_{\rm dm}$ & [$10^6$~\Msun] & $7.5$ & $59$ & $470$ \\ 
$\epsilon^{z=0}_{\rm DM, stars}$ & [kpc] & $0.75$ & $1.48$ & $2.95$ \\
\hline
\end{tabular}
\caption{Primary simulation parameters of the IllustrisTNG runs analysed in this study. The TNG300 simulation is used in the main study, while TNG100 and TNG300-2 are used in Appendix~\ref{app:res} for a resolution study. For a more extensive overview of the parameters of the simulation suite we refer to \citet[][their Table A1]{Nelson+2017}.}
\label{table:sims}
\end{center}
\end{table}

The simulations used in this study are part of the IllustrisTNG project \citep{Marinacci+2017, Naiman+2017, Nelson+2017, Pillepich+2017b, Springel+2017}. These are high-resolution simulations of cosmological structure formation in a representative part of the universe. The primary simulation used for the present analysis has a side length $\sim 300$~comoving~Mpc (TNG300) and follows the formation and evolution of structure governed by the laws of gravity and magnetohydrodynamics from the early universe to redshift zero. We also use a simulation of higher resolution but smaller volume (TNG100) as well as a lower resolution version of the large box (TNG300-2) to test for numerical convergence. The main numerical parameters of these simulation can be found in Table~\ref{table:sims}. For a more detailed list of parameters, see \citet[][their Table A1]{Nelson+2017}.

\subsection{Initial conditions}

The simulations were initialised at redshift~$z=127$ using the \citet{Planck2016} cosmological parameters (i.e.~a matter density $\Omega_m = 0.3089$, baryon density $\Omega_b = 0.0486$, dark energy density $\Omega_\Lambda = 0.6911$, Hubble constant $H_0 = 67.74$~km~s$^{-1}$~Mpc$^{-1}$, power spectrum normalisation $\sigma_8=0.8159$ and a primordial spectral index $n_s = 0.9667$)  and the Zel'dovich approximation for the initial displacement field, which is applied to glass initial conditions \citep{White1994}. The simulations start out with a uniform magnetic seed field with comoving  field strength of $10^{-14}$ Gauss. A suite with several different box sizes at different numerical resolutions were computed as part of IllustrisTNG. 

The highest resolution version of TNG300, which has the largest box size, contains $2500^3$ dark matter particles and the same number of gas cells in the initial conditions. Additionally, we analyse a lower resolution counterpart (TNG300-2) which has the same volume but a factor of $2^3$ reduced particle number, and two times worse spatial resolution with all softenings increased by a factor of $2$. We furthermore analyse the TNG100 simulation from the IllustrisTNG set, which has $2\times 1820^3$ resolution elements, and a factor of $2^3$ higher mass and a factor of $2$ higher spatial resolution than TNG300, but covers only a volume of $\sim 110^3$~Mpc$^3$ and consequently does not contain rare objects such as rich galaxy clusters.

\subsection{Methods}
The simulations are evolved with the \textsc{Arepo} code, i.e. using a finite-volume approach where the equations of ideal magnetohydrodynamics are solved on a quasi-Lagrangian, moving, unstructured mesh \citep{Springel2010,Pakmor+2011,Pakmor+2016}. The divergence constraint of the magnetic field is taken care of by an 8-wave Powell-cleaning scheme \citep{Pakmor+Springel2013}. The gravitational forces are calculated using a tree-particle-mesh method with an operator-split, hierarchical time integration, allowing for efficient calculations of gravitational forces in systems with large dynamic range in time. 

We employ a cooling function using primordial and metal line cooling, a time-dependent ultraviolet background from stars and luminous AGN, prescriptions for star-formation, stellar feedback and metal enrichment, as well as a model for SMBHs, including their formation, growth and feedback effects. The approaches used in the IllustrisTNG simulations (with identical parameter choices), as well as the impact of variations of model parameters, are presented in two separate method papers, \citet{Pillepich+2017} for the stellar feedback, enrichment and the low mass end of the galaxy stellar mass function (GSMF), and \citet{Weinberger+2017} for the AGN feedback model and the high mass end of the GSMF. Here, we only briefly summarize the aspects of the model that are most relevant to this study.

\subsection{Modelling of supermassive black holes}
\label{sec:BH_model}

\begin{figure*}
  \includegraphics{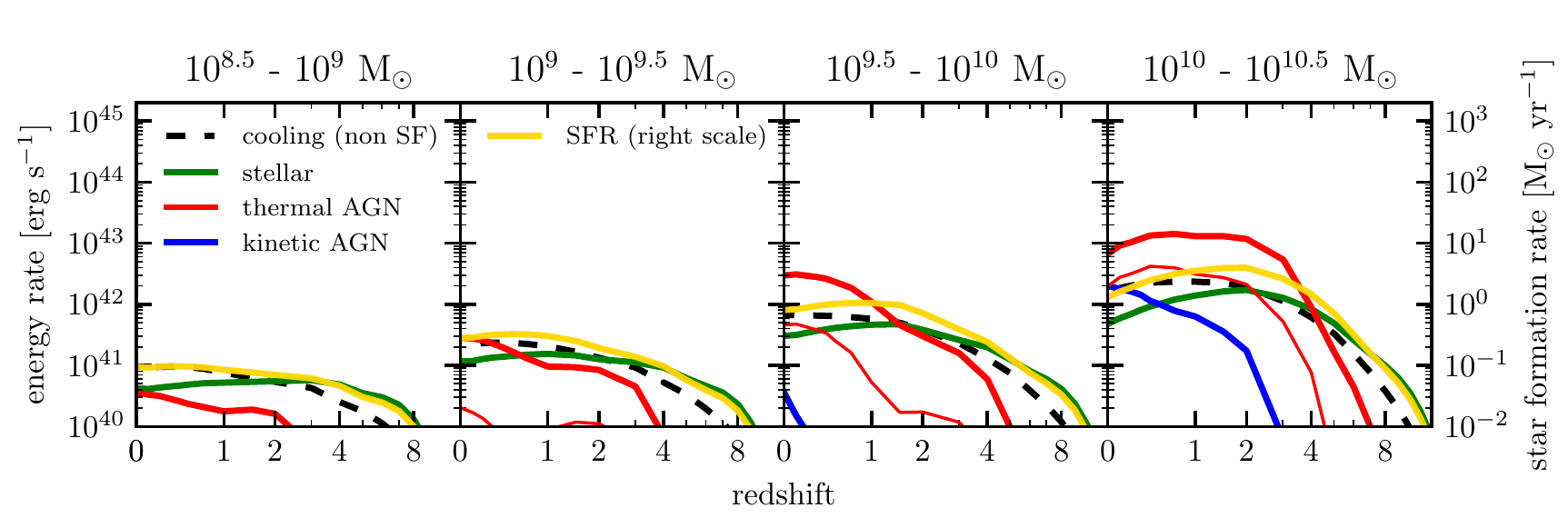}
  \includegraphics{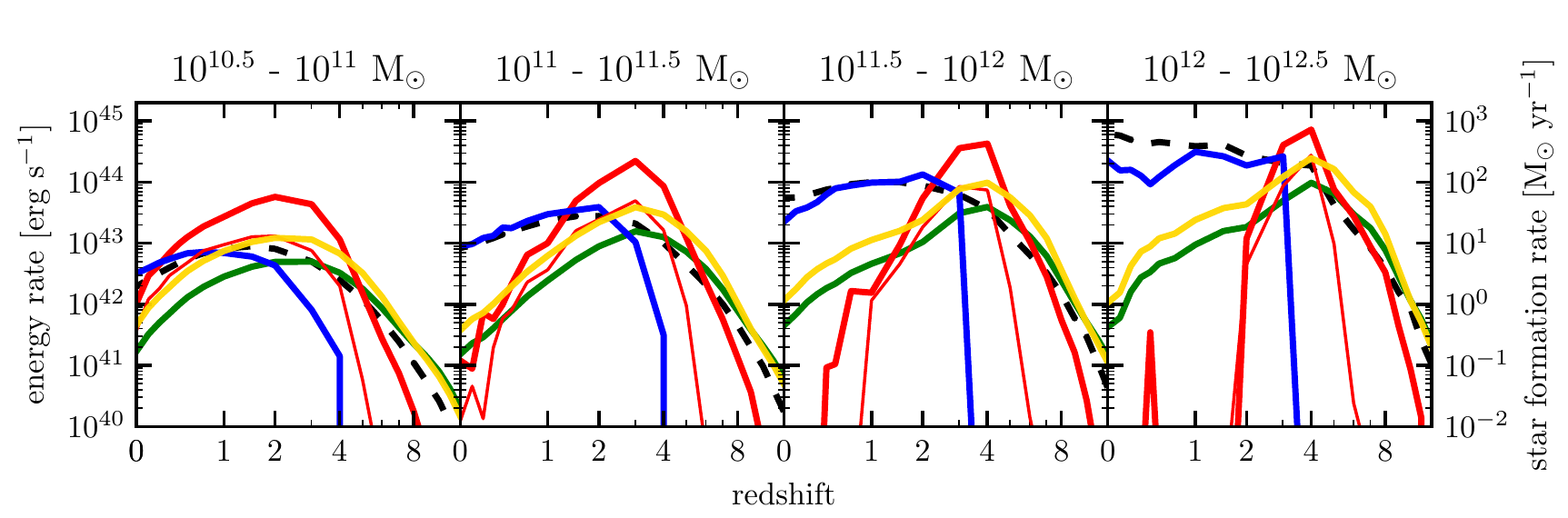}
  \caption{Average instantaneous feedback energy rates (thermal AGN feedback in red, kinetic AGN feedback in blue, stellar feedback in green) and non-star-forming gas cooling rate (dashed black) within each galaxy as a function of redshift for central galaxies with different redshift $z=0$ stellar mass. The yellow line indicates the star formation rate in M$_\odot$~yr$^{-1}$ (right scale). Independent of galaxy mass, stellar feedback always dominates at high redshift, followed by thermal AGN feedback. In massive halos, the kinetic AGN feedback takes over at late times, approximately compensating the cooling losses and keeping the star formation rate low. The thin line indicates the effective thermal AGN feedback energy rate (see text), which is substantially lower than the nominal value, in particular in low-mass systems. Note that there is an additional heating channel via gravitational infall of gas, which we do not account for here.}
  \label{fig:FeedbackEnergy_vs_redshift}
\end{figure*}

We identify friend of friends (FOF) groups on the fly during the simulation \citep{Springel+2001b}.
A SMBH with mass $1.18 \times 10^{6}$~\Msun is seeded whenever a FOF halo exceeds a mass of $7.38 \times 10^{10}$~\Msun and does not yet contain a SMBH. These black holes are then accreting according to an Eddington-limited Bondi accretion rate

\begin{align}
 \dot{M}_\text{Bondi} &= \frac{4 \pi G^2 M_\text{BH}^2 \rho}{c_s^3}, \\
 \dot{M}_\text{Edd} &= \frac{4 \pi G M_\text{BH} m_p}{\epsilon_r \sigma_{\rm T}} c, \\
 \dot{M} &= \min\left( \dot{M}_\text{Bondi} , \dot{M}_\text{Edd}  \right),
\end{align}
where $G$ is the gravitational constant, $M_\text{BH}$ the mass of the black hole, $\rho$ the kernel-weighted ambient density around the SMBH, $c_s$ the kernel-weighted ambient sound speed including the magnetic signal propagation speed, $m_p$ the proton mass, $c$ the speed of light, $\epsilon_r = 0.2$ the black hole radiative efficiency and $\sigma_{\rm T}$ the Thompson cross section. We emphasize that, unlike previous work \citep[e.g.][]{Springel+2005}, this model does not use an artificial boost factor in the accretion rate. We enforce the SMBH to be located at the potential minimum of its host halos at every global integration timestep, and assume a prompt merging of SMBH binaries. Note that the SMBH merger rates, which will be discussed in this paper, are not affected by our neglect of delay times in the merging process (assuming all pairs eventually merge).

Feedback from SMBHs is injected in two different channels, where the dividing line is the accretion rate in units of the Eddington accretion limit. Whenever the Eddington ratio exceeds a black hole mass dependent threshold of
\begin{align}
  \chi = \min\left[ 0.002 \left(\frac{M_\text{BH}}{10^8 \,\text{M}_\odot} \right)^2, 0.1\right],
\end{align}
 the feedback energy is injected continuously as thermal energy (`thermal mode') into the surroundings of the black holes with a rate of $\dot{E}_{\rm therm} = 0.02\, \dot{M} \,c^2$, while for lower accretion rates the feedback energy is injected into the surroundings as pure kinetic feedback (`kinetic mode') in a pulsed, directed fashion with the rate $\dot{E}_{\rm kin} = \epsilon_{f,\text{kin}}\, \dot{M} \,c^2$, where
\begin{align}
  \epsilon_{f,\text{kin}} = \min\left( \frac{\rho}{0.05 \,\rho_\text{SFthresh}} , 0.2\right),
\label{eqnsurroundfac}
\end{align}
and $\rho_\text{SFthresh}$ is the star formation threshold density. 
The factor (\ref{eqnsurroundfac}) means that we assume that at low environmental densities, the coupling of the AGN feedback energy to the surroundings becomes weak. We employ a similar scaling in the thermal mode, where we use the approach of \citet{Vogelsberger+2013} and reduce the accretion rate by a factor of $\left(P_{\rm ext}/P_{\rm ref}\right)^2$ whenever $P_{\rm ext} < P_{\rm ref}$. $P_{\rm ext}$ is the kernel weighted pressure of the gas surrounding the black hole and $P_{\rm ref}$ is a reference pressure defined in \citet[][their equation 23]{Vogelsberger+2013}.

\subsection{Feedback energetics}

The stellar feedback parametrisation is described in \citet[][their section 2.3.2]{Pillepich+2017}. The energy of the stellar feedback is given by their equation~3, which can be rewritten as

\begin{align}
\dot{E}_{\rm stellar} &= 3.41 \times 10^{41} {\rm erg\,s^{-1}} \left(\frac{\rm SFR}{\rm M_\odot\,yr^{-1}}\right) \, f(Z) ,
\label{eq:stellarFeedback}
\\
f(Z) &= 1 + \frac{3}{1+(Z/0.002)^{2}},
\end{align}
where $Z$ is the metallicity (metal mass fraction) and ${\rm SFR}$ the star formation rate in a gas cell. With this parameterisation, $f(Z)$ is usually close to unity, but for very low metallicity it can increase up to a value of $4$ to account for reduced cooling losses in this regime, which happens preferentially at high redshift for low-mass systems.

The maximum AGN feedback energy rate (i.e. not limited due to the lower efficiencies at low surrounding densities) can be written as
\begin{align}
  \dot{E}_{\rm thermal\, AGN} &= {5.66 \times 10^{42} {\rm erg\,s}^{-1}}\frac{ \dot{M}}{5\times 10^{-3}\,{\rm M}_\odot\,{\rm yr}^{-1}}, \label{eq:thermalAGNfeedback}\\
  \dot{E}_{\rm max.\, kinetic\, AGN} &= {5.66 \times 10^{43} {\rm erg\,s}^{-1}}\frac{ \dot{M}}{5\times 10^{-3}\,{\rm M}_\odot\,{\rm yr}^{-1}}, \label{eq:kineticAGNfeedback}
\end{align}
where the different reference value for gas accretion compared to the star formation rate in equation~(\ref{eq:stellarFeedback}) is inspired by the black hole mass--stellar bulge mass relation \citep{Kormendy+Ho2013}. Comparing equations~(\ref{eq:stellarFeedback}) and (\ref{eq:thermalAGNfeedback}), it becomes clear that a system which is located on the black hole mass -- stellar bulge mass relation will have experienced significantly more energy injection from AGN feedback than from stellar feedback (provided the contribution from black hole seeds to the SMBH mass is subdominant). 

However, this does not automatically imply that AGN feedback is the dominant feedback channel in these galaxies, i.e.~is mainly responsible for regulating the star formation rate in these systems. The feedback efficiency depends on the precise way the feedback energy is injected into the surrounding medium \citep[e.g.][]{Rosdahl+2017, Smith+2017}. In the employed stellar feedback model, cells are attributed a probability (proportional to the star formation rate) to launch a `wind particle' with a given velocity that is temporarily hydrodynamically decoupled. The particles recouple to the gas as soon as they reach a cell with a density lower than $0.05$ times the star formation threshold \citep{Springel+Hernquist2003, Pillepich+2017}. This means that the coupling of the stellar wind feedback to the gas is slightly non-local and hence different from the AGN feedback, which is directly injected to the surrounding gas cells. Therefore, relating the energetics of the different feedback channels to their overall importance for galaxy formation requires a careful analysis of the simulation results and is not possible from simple analytic considerations alone.

What is clear, however, is that the kinetic AGN feedback channel is significantly more efficient than the thermal AGN feedback mode, which is one of the key features of the employed AGN feedback implementation. Besides the higher efficiency parameter in the kinetic mode (equations~\ref{eq:thermalAGNfeedback} and \ref{eq:kineticAGNfeedback}), this is due to the fact that a pulsed injection of feedback energy (as used in the kinetic mode) heats up the affected gas to higher temperatures and consequently reduces the cooling losses compared to a continuous injection (used in the thermal mode).

\subsection{Definition of quenched galaxies} 

Throughout this paper, we analyse the main, central halos (excluding satellites) of the TNG300 simulation, the largest volume simulation box of the IllustrisTNG project. We mainly focus on halos with redshift $z=0$ stellar masses (measured within twice the stellar half mass radius) larger than $10^{10.5}\,{\rm M}_\odot$, and use the relation
\begin{align}
 \log {\rm SFR}_{\rm SFMS} = -7.4485 + \log M_* \times 0.7575
 \label{eq:SFMS}
\end{align}
adopted from \citep{Ellison+2015}, where SFR is the star formation rate in M$_\odot$~yr$^{-1}$ and $M_*$ the stellar mass within twice the stellar half mass radius in M$_\odot$, to define the star forming main sequence (SFMS) at redshift $z=0$. We define quenched galaxies as systems with an instantaneous star formation rate (measured from gas cells within the same radius) of at least $1$~dex below SFR$_{\rm SFMS}$, independent of their redshift. We consider all systems above this cut as star-forming. 
The use of the instantaneous star formation rate measured from the star-forming gas cells \citep[given as an output of the model of][]{Springel+Hernquist2003} is helpful for this study, as it just depends on the present state of the gas in the galaxy and not on its history. However, we also emphasize that this instantaneous star formation rate is not an observable quantity and it can vary on rather short timescales, which increases the scatter in its distribution function. Therefore, we leave a detailed comparison to the numerous observations in this area \citep[e.g.][]{Noeske+2007,Rodighiero+2011,Wuyts+2011,Whitaker+2012,Speagle+2014,Renzini+Peng2015,Schreiber+2015,Tasca+2015} to future work, and use the star formation rate in this paper only as a proxy for classifying a galaxy to be either in a quenched or star-forming state. 

\section{The galaxy population}
\label{sec:GalaxyPopulation}

Because of its large volume and high number of resolution elements, the TNG300 simulation contains an unprecedented number of resolved galaxies in a single hydrodynamic simulation, ranging from isolated Milky Way-sized galaxies to massive brightest cluster galaxies. In particular, there are $19090$ redshift $z=0$ galaxies with a stellar mass larger than $10^{10.5}\,{\rm M}_\odot$. In this paper, we focus mainly on these massive galaxies that host a supermassive black hole and are significantly affected by its feedback energy. In particular, we want to answer the question how the massive end of the galaxy population was driven off the SFMS and became quiescent, and how this relates to observables of black hole activity.

\subsection{Feedback at different galaxy masses}

To understand the behaviour of galaxies of different masses in the simulation, we first show the energetics of the gas phase of these systems. To this end, we select central galaxies in different redshift $z=0$ stellar mass bins and trace back their main progenitor. Throughout this paper, we define the stellar mass as the mass within twice the stellar half-mass radius. As in previous work \citep[e.g.][]{Nelson+2017, Genel+2017}, we use the merger-tree algorithm described in \citet{Rodriguez-Gomez+2015} to track the central galaxies identified by the \textsc{Subfind} algorithm \citep{Springel+2001b} to high redshift. In the following, we refer to a `galaxy' as the subhalo identified by \textsc{Subfind}. We calculate the instantaneous cooling rate of all non-star-forming cells in the galaxy\footnote{The thermodynamic state of all star forming cells is described by an `effective equation of state' model \citep{Springel+Hernquist2003}, which makes it difficult to define an unambiguous cooling rate for them.} as well as the instantaneous feedback energy rates of stellar feedback, thermal AGN feedback and kinetic AGN feedback in the respective subhalos. The stellar feedback is calculated via the star formation rate and gas metallicity on a cell-by-cell basis using equation~(\ref{eq:stellarFeedback}), and the AGN feedback energy by using the black hole accretion rate and applying the appropriate formulae for either thermal or kinetic feedback given in Section~\ref{sec:BH_model}, using the most massive SMBH in the subhalo (using all black holes in a subhalo instead does not change the results). Note that an individual SMBH can only be in either of the two modes, which means that the average, i.e.~the sum over all active SMBH divided by the total number of galaxies in this stellar mass bin, can suddenly drop to zero whenever the SMBHs in these subhalos are not in the corresponding mode. 

We show the redshift evolution of the average energy rates and the average star formation rate in the corresponding galaxies in Figure~\ref{fig:FeedbackEnergy_vs_redshift}. For galaxies of all masses, stellar feedback dominates at high redshift. For the lowest mass bin shown here ($10^{8.5}-10^{9}$~M$_\odot$), stellar feedback (green line) stays the dominant feedback channel until redshift $z=0$, while in the most massive galaxies, the thermal AGN feedback (red line) becomes dominant over stellar feedback already at $z=6$. Note that the stellar feedback becomes sub-dominant in part because the AGN feedback reduces star-formation, so one needs to be aware that the AGN feedback indirectly reduces the stellar feedback energy. At low redshifts, the AGN feedback energy dominates for all but the least massive galaxies shown here. We emphasise, however, that this does not necessarily mean that it is the most important feedback channel in these systems. In particular, in our model, the stellar feedback energy, by construction, only couples to non-star forming gas which has comparably low density and consequently low cooling losses, while the thermal AGN feedback is continuously injected around the SMBH. As the region around the SMBH can contain very dense, star forming gas, large fractions of the injected feedback energy can be radiated away immediately. We expect the kinetic AGN feedback to be less affected by this effect due to the kinetic, pulsed injection of momentum, and therefore to be more efficient than the thermal mode at equal energy rates. 

The cooling rate in the non-star-forming phase (dashed black line) is of the same order as the feedback energy of stellar and kinetic AGN feedback. The thermal AGN feedback energy can significantly exceed this rate without having any dramatic effect on the star formation rate (yellow line), indicating that large amounts of this energy are lost in the star forming gas phase. To illustrate this point further, the thin red lines in Figure~\ref{fig:FeedbackEnergy_vs_redshift} show the effective thermal AGN feedback energy rate. This quantity is calculated by summing up the contribution of thermal AGN feedback of each individual gas cell in the surroundings of a SMBH.
We then subtract the cooling losses if the gas cell is in the star-forming regime. If the cooling losses exceed the feedback energy in a particular cell, its contribution to the feedback energy neglected. This means that we only take the energy injection in gas cells that are not in the star-forming regime in the subsequent timestep into account. The energy difference between the nominal and the effective thermal AGN feedback is de facto never injected, as the respective gas cells stay on the effective equation of state, thus have a pre-defined pressure given their density. In this way, we obtain the actual thermal feedback energy that is not immediately lost to cooling, which is a more realistic estimate of the feedback energy from this channel.

Thus, thermal AGN feedback energy only becomes an important feedback channel in galaxies with stellar masses of around $10^{10}\,{\rm M}_\text{\sun}$. Wherever the kinetic AGN feedback takes over, i.e.~in galaxies more massive than $10^{10.5}\,{\rm M}_\text{\sun}$ at low redshift, the star formation rate is significantly reduced and the shape of the star formation rate curve no longer follows the cooling curve, but is significantly suppressed. We now investigate how this suppression in star formation rate impacts the specific star formation rate -- stellar mass diagram and we examine on a galaxy-by-galaxy basis what triggers the quenching initially.

\subsection{Star formation in galaxies}

\begin{figure*}
  \includegraphics{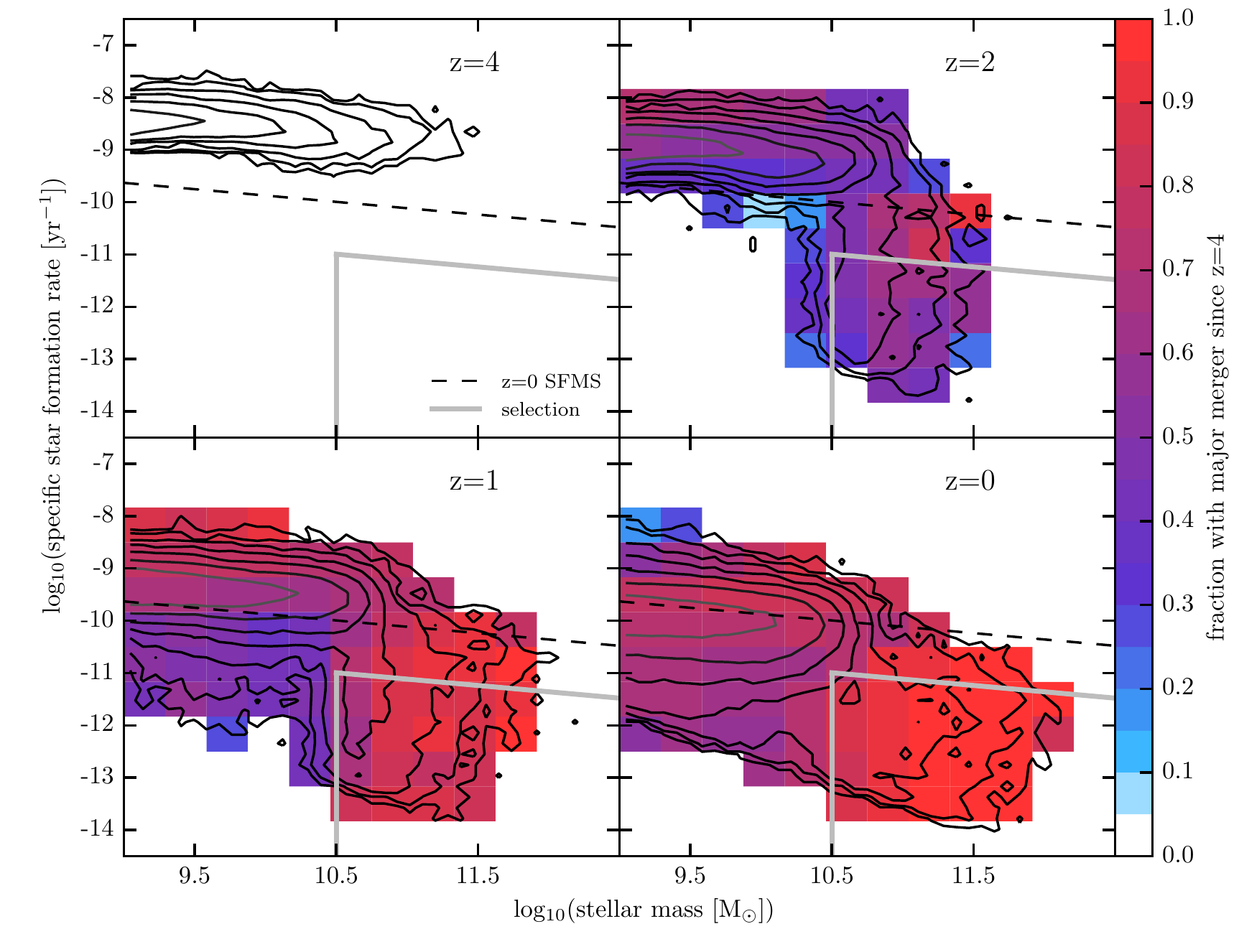}
  \caption{Specific star formation rate vs stellar mass in the TNG300 simulation. The contours indicate number density (the outermost contour encloses a density of $10^{-5.5}$ comoving Mpc$^{-3}$ dex$^{-2}$, the other contours show a density increase by $0.5$~dex each) in this plane, the colours indicate the fraction of systems in the corresponding bin which have had a major merger (subhalo stellar mass ratio >1:4) since $z=4$, i.e. within the past $\sim12.3$ Gyr. The dashed line shows the star forming main sequence at $z=0$, adopted from \citet{Ellison+2015}. Most massive, quenched galaxies, have undergone at least one major merger by $z=0$. However, this is not the case for quenched galaxies at higher redshift, where, e.g. at $z=2$, only about half of the quenched galaxies have undergone at least one major merger. Therefore, major mergers cannot be the sole reason for quenching. The grey lines indicate the selection for quenched massive central galaxies.}
  \label{fig:SFR_Mstar}
\end{figure*}

\begin{figure}
  \includegraphics{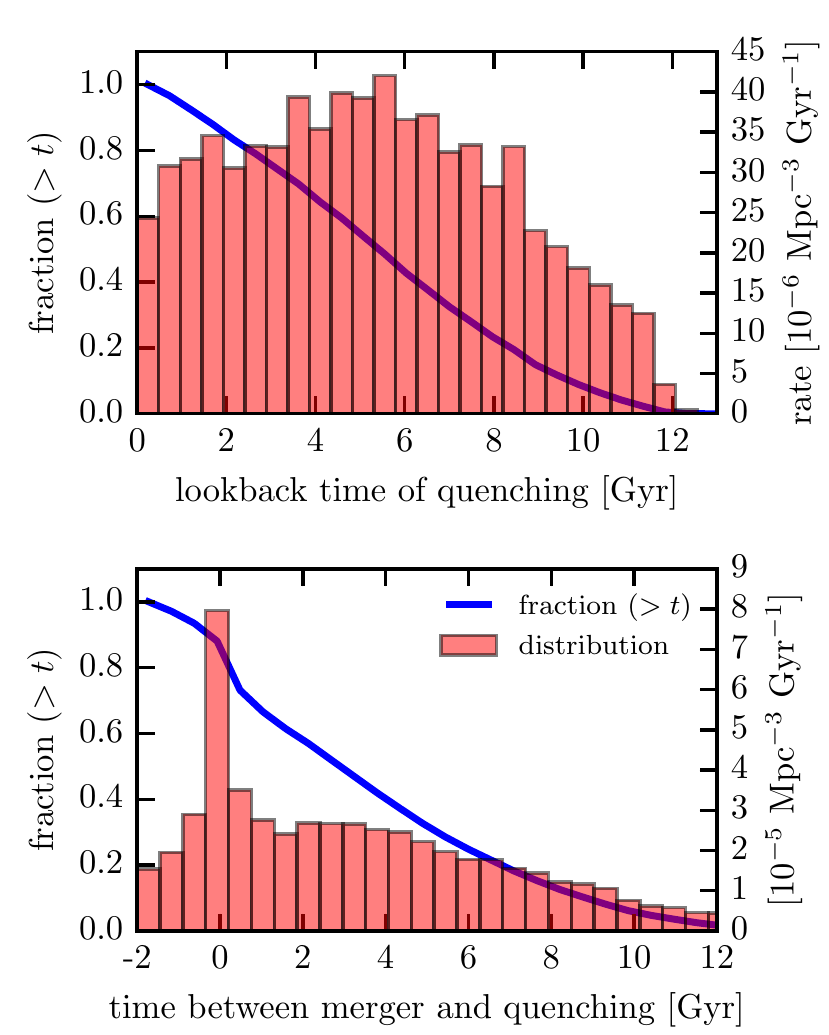}  
  \caption{{\it Top panel:} Distribution of the lookback time when the (last) quenching happens for more than $17000$ quenched central galaxies with a redshift $z=0$ stellar mass larger than $ 10^{10.5}$~M$_\odot$ and a star formation rate at least $1$~dex below the star forming main sequence. {\it Bottom panel:} Distribution of time between last quenching and last major merger prior to quenching (see text for precise definition). The scale of the histogram is shown on the right axis, while the cumulative distribution function is shown on the left axis.}
  \label{fig:delayTime}
\end{figure}

Figure~\ref{fig:SFR_Mstar} shows the specific star formation rate (defined as the star formation rate divided by the stellar mass) within twice the stellar half mass radius as a function of stellar mass for all central galaxies in the simulation. At redshift $z=0$, there is a well-defined star forming population up to stellar masses of around $5\times 10^{10}\,{\rm M}_\odot$, following the dashed line which denotes the observed SFMS from \cite{Ellison+2015}. Starting at $3\times 10^{10}\,{\rm M}_\odot$, there is a significant population of galaxies off the SFMS. The different panels show the same range of specific star formation rate vs stellar mass at redshifts $z=4$, $2$, $1$ and $0$. At redshift $z=4$, the bulk of the galaxy population at all masses is star-forming. At $z=2$, a significant fraction of the high-mass galaxies have already quenched, while the main sequence still shows a very high level of specific star formation rate. This changes towards $z=1$, where the lower mass population also has a larger low specific star formation rate tail, which is even more pronounced at $z=0$.  

The colour coding in this plot shows the fraction of systems that experienced a major merger ( $>1:4$ in stellar mass ratio\footnote{We measure the merger masses as the stellar mass of the galaxy at the time of maximum mass of the low-mass progenitor \citep[see][]{Rodriguez-Gomez+2015}.}) since redshift $z=4$. For the $z=0$ massive galaxies, most systems experienced at least one major merger. Looking at the onset of the population of quenched galaxies at $z=2$, however, a significant fraction of the quenched galaxies have not undergone such a major merger, and no significant enhancement of mergers can be seen in this population relative to the SFMS galaxies. This means, in particular, that a large fraction of galaxies that are quenched by $z=2$ did not undergo a major merger since $z=4$, which indicates that a scenario in which a system needs to undergo a gas-rich major merger leading to a starburst in order to trigger subsequent quenching by AGN activity \citep{DiMatteo+2005, Springel+2005, Hopkins+2008} does not apply to all quenched galaxies in IllustrisTNG (but it might still do to a sub-population).

To investigate this further, we select all central galaxies more massive than $10^{10.5}\,{\rm M}_\odot$ in stars and that are at least $1$~dex below the SFMS (equation~\ref{eq:SFMS}) at redshift $z=0$. Because of the large volume of the simulation, we find more than $17000$ such systems. After tracing the main progenitor branch back in time, we define the (last) time of quenching as the snapshot after they were located above this selection threshold for the last time\footnote{The precise value of the threshold does not change the conclusions drawn here.}. The distribution function of the quenching times is shown in the top panel of Figure~\ref{fig:delayTime}. We then identify major mergers prior to the time of quenching and measure the time between the last merger and the quenching. The resulting distribution function of this time difference is shown in Figure~\ref{fig:delayTime}, bottom panel. We note that we perform this analysis in post-processing on $100$ snapshots which are roughly equally spaced in scalefactor, which gives a time-resolution of around $200$~Myr at low redshift. In practice, we also include mergers that, according to the merger-tree algorithm, happen up to $2$~Gyr after quenching, as these not yet merged galaxies might have tidal interactions with the host, which can cause AGN activity. Increasing this time does not change the result. There is an excess of systems that had experienced a recent merger prior to quenching, but, more significantly, a tail which extends all the way to $12$~Gyrs. This means that for the majority of the quenched galaxies in IllustrisTNG, we cannot relate their quenching to a particular major merger event. 

\begin{figure}
  \includegraphics{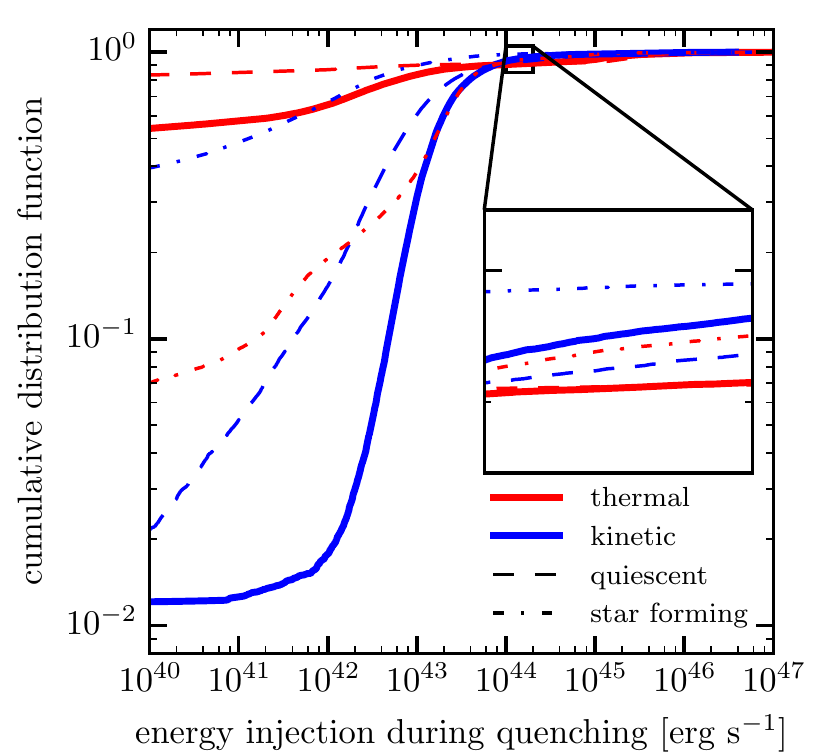}
  \caption{Cumulative distribution function of the average energy injected between the last snapshot when galaxies are found above $0.1$ times the SFMS and the first time below this threshold. We consider the same galaxies here as in Figure~\ref{fig:delayTime}. Note that the spacing between two snapshot outputs is around $200$~Myr. The dashed and dash-dotted lines measure the same quantity for a sample of quiescent and star forming galaxies, respectively, with the same distribution in redshift as the quenching events.}
  \label{fig:EgyAtQuenching}
\end{figure}

To determine what happens during quenching, we use the identified quenching events and evaluate the time-averaged AGN feedback energy injection during the period in which the galaxies transition to the quenched population. Technically, we derive this quantity by using the cumulative energy injected in the thermal and in the kinetic AGN mode (these are kept track of in the simulation and are part of the output for every SMBH) from all SMBHs in a given galaxy at the snapshot directly after quenching. We then use the SMBH merger tree to identify all progenitors of those SMBHs in the last snapshot where the host galaxy was still star forming, i.e.~the star formation rate was larger than 0.1 times the corresponding observed $z=0$ SFMS value (equation~\ref{eq:SFMS}), and subtract the cumulative energy up to this snapshot from the final one. We then divide the remaining energy differences by the time elapsed between the two snapshots (typically around $200$ Myr), and therefore get an average feedback energy rate. 

We plot the cumulative distribution function (CDF) of this average feedback rate, i.e. the fraction of systems with a feedback energy lower than the given value, both for the thermal (solid red line) and kinetic (solid blue line) modes, in Figure~\ref{fig:EgyAtQuenching}. From this figure, it becomes clear that more than $90 \%$ had an average kinetic AGN feedback energy larger than $10^{42.5}\,{\rm erg\,s}^{-1}$, whereas more than $50 \%$ of the galaxies had an average thermal feedback energy of less than $10^{40}\,{\rm erg \, s}^{-1}$. We speculate that the $1-2 \%$ of galaxies without significant kinetic feedback are redshift $z=0$ central galaxies that were satellites at quenching, but leave an explicit confirmation of this to future work. 

Additionally, we select a sample of star forming and a sample of quenched galaxies with the same redshift $z=0$ mass cut and the same redshift distribution, and compare the feedback energy during star forming (dash-dotted) and quiescent (dashed) phases. The comparison shows that galaxies that are quenching have a higher kinetic AGN feedback rate than quiescent systems (except for a very small sub-population, which likely originates in an implicit mass-selection effect when selecting for quiescent galaxies), and that kinetic AGN feedback is energetically unimportant for more than half of the star-forming systems even in these high mass systems ($>10^{10.5}\,{\rm M}_\text{\sun}$). We therefore conclude that kinetic mode AGN feedback causes the quenching (as well as quiescence) of massive central galaxies in IllustrisTNG.

By construction, it becomes easier for the AGN to enter the kinetic mode once the SMBHs are massive  (around $10^8\,{\rm M}_\odot$) and have a low accretion rate relative to the Eddington limit \citep{Weinberger+2017}. To study the connection between quenching and the supermassive black hole mass, we plot the star formation efficiency (SFE), defined as the star formation rate divided by the gas mass in twice the stellar half mass radius, as a function of black hole mass. We bin the distribution, colour-coded by average stellar over black hole mass, in the top panel of Figure~\ref{fig:Mbh_Mstar}. In case there is more than one black hole in the galaxy, we use the mass of the most massive one. The grey line indicates the average star formation efficiency. 

There is a sharp increase in the SFE with stellar mass for galaxies with black holes with a mass close to the seed mass, as well as a steep drop above $\sim 2 \times 10^8\,{\rm M}_\odot$. At these high masses, there is also a significant number of galaxies with zero star formation rate which do not enter this plot. We note, however, that there are also individual systems that have a SFE of around $10^{-10}\,{\rm yr\,}^{-1}$. Apart from the highest SFE values at black hole masses of around $10^{7.5}\,{\rm M}_\odot$, which seem to have a particularly low-mass black hole for their stellar mass, there is no significant trend of SFE with stellar mass in this plot. The ratio of stellar mass over black hole mass has a noticeable drop at around $10^{7.5}\,{\rm M}_\odot$, with significantly under-massive SMBH at lower black hole masses due to a delayed growth of SMBHs after seeding, and a roughly constant stellar mass to black hole mass ratio of $\sim200$ at higher SMBH masses, which will be discussed in the next section. 

Looking at the black hole mass -- stellar mass plane in Figure~\ref{fig:Mbh_Mstar} (bottom panel, colour-coded by the average of the star-formation efficiency), it becomes clear that the change in star formation efficiency at black hole masses of a few times $10^8\,{\rm M}_\odot$ is even more significant for systems with over-massive black holes and stellar masses around $10^{10.5}\,{\rm M}_\odot$, manifesting itself in a population with zero mean star formation rate (within the contours but no colour-coding).

\section{The connection to the SMBH population}
\label{sec:SMBH}

\begin{figure}
  \includegraphics{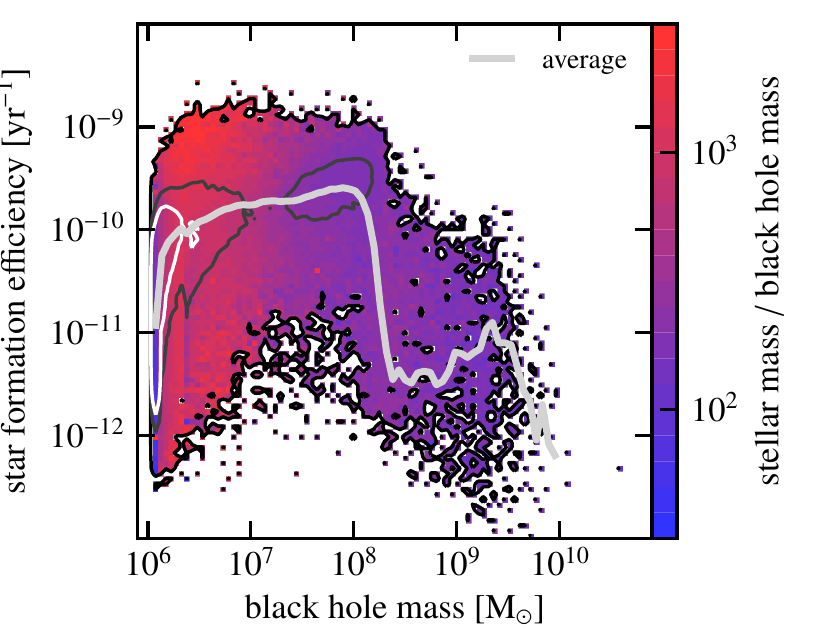}
  \includegraphics{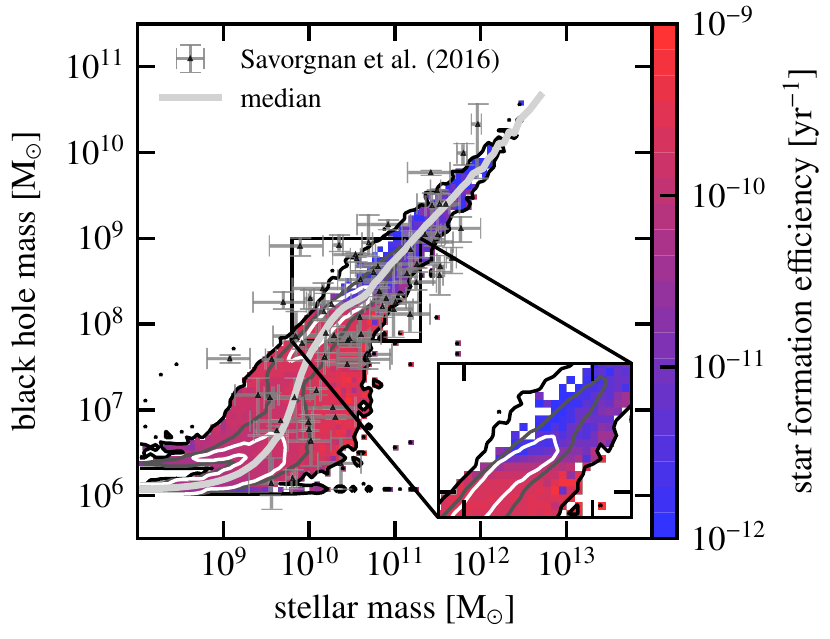}
  \caption{{\it Top panel:} 2d histogram of star formation efficiency, defined as the star formation rate divided by the gas mass within twice the stellar half mass radius, vs black hole mass, colour coded by average stellar over black hole mass. The solid grey line shows the average. {\it Bottom panel:} Black hole masses vs stellar mass colour coded by the star formation efficiency. The solid grey line shows the median. The symbols with errorbars are observational data taken from \citet{Savorgnan+2016}.}
  \label{fig:Mbh_Mstar}
\end{figure}

In general, the black hole mass -- stellar mass relation (Figure~\ref{fig:Mbh_Mstar}, bottom panel) agrees well with observational data from \citet{Savorgnan+2016}, in the sense that the observational data could be drawn as a subset of the simulated objects. We note, however, that due to a lack of resolution we do not perform a decomposition of each galaxy to derive a mass for the bulge component in our simulation data. This aspect, as well as the resolution dependence of both stellar \citep[e.g.][their Appendix B]{Weinberger+2017} and black hole mass (Appendix~\ref{app:res}), leads to some uncertainties in the theoretical prediction. The scatter in the simulation prediction is smaller than in the observational sample, which is a generic feature of many simulation models \citep[e.g.][their Section 3.3 and references therein]{Volonteri+2016}. To quantify our comparison of the scatter, we added Gaussian random noise with a standard deviation equal to the average measurement errors of \citet{Savorgnan+2016} to the simulation data (all in log-space), and measured the root mean square distance from the mean of the logarithm of the black hole mass over stellar mass fraction in stellar mass bins of $1$~dex width (ranging from $10^{9}\,{\rm M}_\odot$ to $10^{12}\,{\rm M}_\odot$). The resulting scatter is around $0.4$~dex at $10^{9.5}\,{\rm M}_\odot$ and $0.3$~dex at $10^{11.5}\,{\rm M}_\odot$. This is significantly smaller than in the observations ($0.7$~dex and $0.4$~dex, for $10^{9.5}\,{\rm M}_\odot$ and $10^{11.5}\,{\rm M}_\odot$, respectively). On top of this, the fact that the observational sample is highly biased means that the discrepancy is probably even more severe, because it is hard to imagine how the complete sample could have a smaller dispersion than a specific sub-selection \citep[see e.g.][]{Reines+Volonteri2015}.

Another interesting prediction of the simulation is that there are no significantly over-massive black holes (more than $\sim 0.5$~dex above the median). A few such systems do exist in the simulation, but they are all satellite galaxies, which are excluded from the plot shown here \citep[see also][]{Barber+2016, Volonteri+2016}. We now focus on the question why this is the case, or conversely, why there are no massive central galaxies that have over-massive black holes. It turns out that this question is intimately linked to how black holes grow and how this relation gets established in the first place.
 
\subsection{The black hole mass growth}

\begin{figure}
  \includegraphics{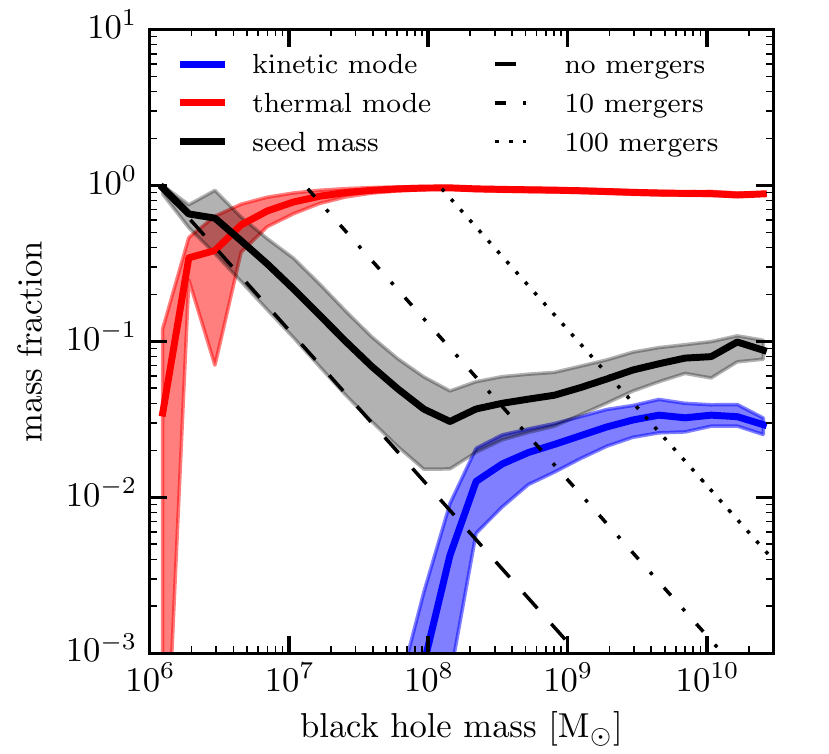}
  \caption{Fraction of mass growth of all progenitors of a redshift zero SMBH via different channels as a function of their redshift $z=0$ mass. The solid lines show the averages, the shaded regions the $10$th and $90$th percentiles. Red and blue lines denote the growth in thermal and kinetic mode, respectively, and the black line indicates the contribution of seed masses to the final black hole mass. Note that the three lines add up to unity by construction. The intersection of the black lines with the dashed, dash-dotted and dotted lines gives the number of mergers the black hole and all its progenitors experienced. The mass growth is dominated by the thermal mode, with a $10$ percent contribution from black hole seeds. Mass growth in the kinetic mode never contributes more than a few percent of a SMBH's mass. }
  \label{fig:FractionMassGrowth}
\end{figure}

\begin{figure*}
  \includegraphics{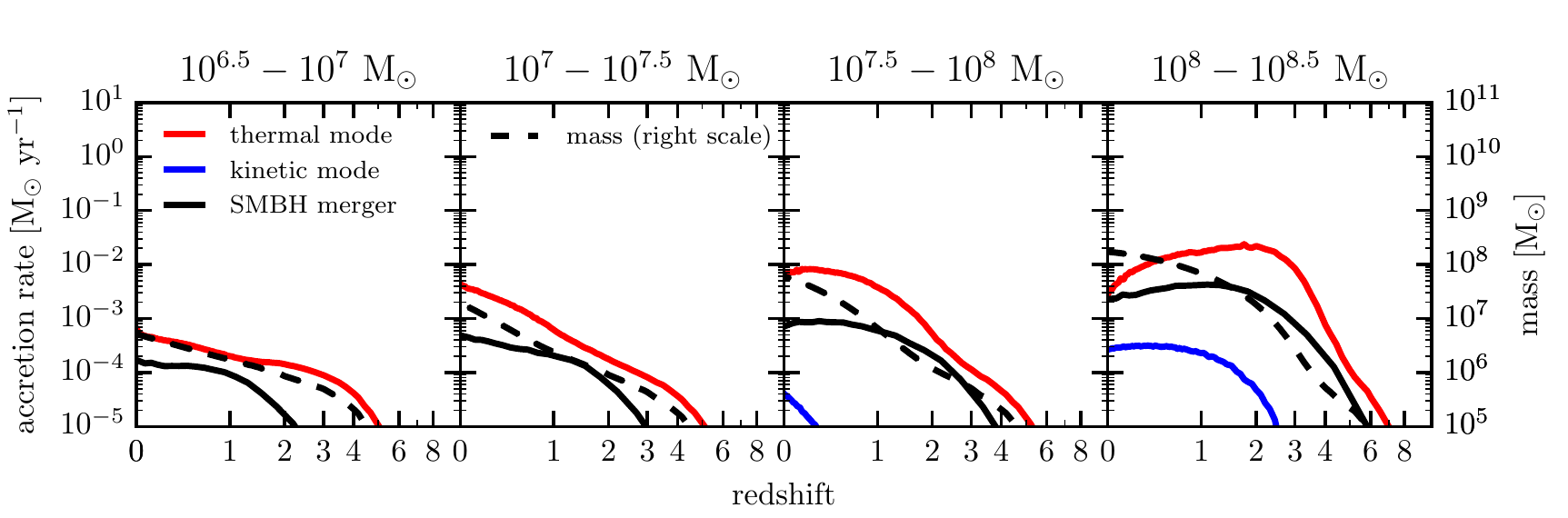}
  \includegraphics{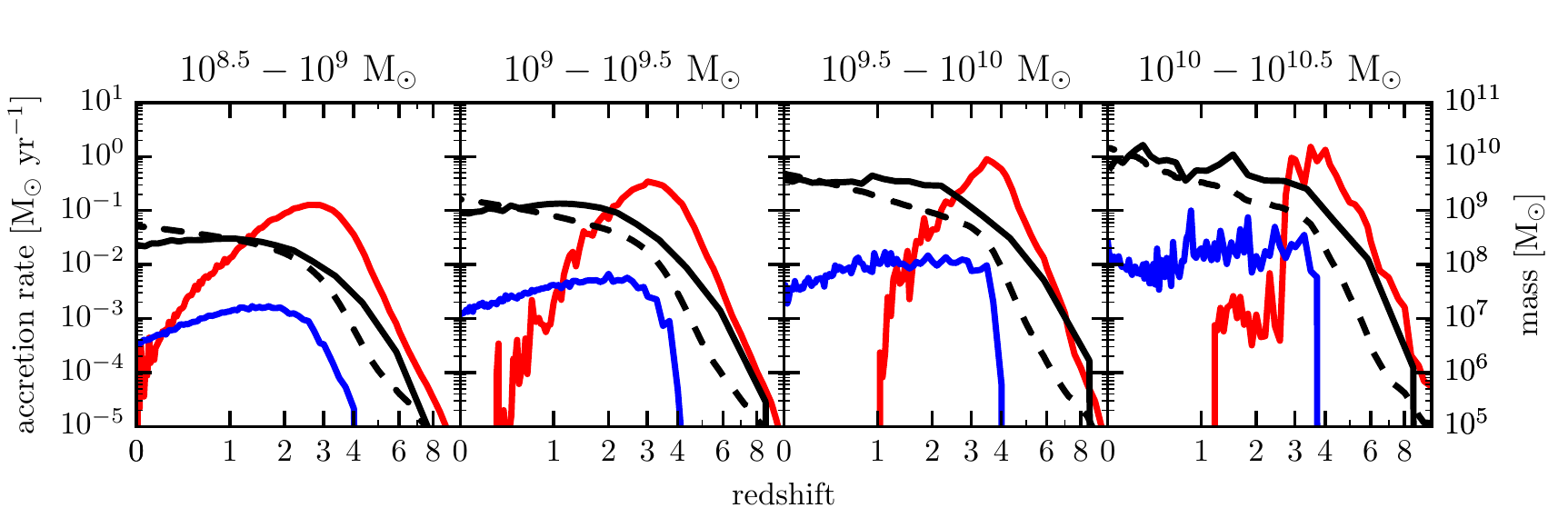}
  \caption{Mean mass growth history of SMBHs of different final masses, split up by gas accretion in thermal (red) and kinetic (blue) feedback mode, as well as via mergers with lower-mass SMBHs (solid black). The more massive progenitor in a SMBH merger defines the main branch of a SMBH, while the less massive SMBH contributes to the merger (solid black) line. The dashed black lines indicate the average mass of the SMBHs at a given redshift. Low mass black holes grow via accretion in the thermal mode, while high mass black holes have a rapid accretion phase at high redshift, until they reach a mass of $\sim10^{8.5}$~M$_\odot$, and build up most of their mass via mergers at later times. }
  \label{fig:MassGrowthRate}
\end{figure*}

Figure~\ref{fig:FractionMassGrowth} shows the overall contribution of accretion in thermal (red) and kinetic mode (blue), as well as the contribution of SMBH seeds to the final black hole mass. The latter is also a measure for the number of mergers of the SMBH and all its progenitors, which increases from an average of $1$ for black holes less massive than $10^{7.5}$~M$_\odot$ to more than $1000$ for the most massive black holes in the simulation ($>10^{10}$~M$_\odot$). Consequently, the seed mass contribution of the most massive black holes reaches about $10\%$. Apart from the least massive black holes, accretion in the thermal mode always dominates the mass growth of black holes, while growth in the kinetic mode is completely subdominant at all masses.

Note however that Figure~\ref{fig:FractionMassGrowth} does not distinguish whether the mass growth in the thermal mode was taking place in-situ or by merging with lower mass SMBHs, which themselves grew via accretion in the thermal mode (the number of progenitors for high-mass black holes shows that the latter scenario is plausible for them). To investigate this in detail, we plot the instantaneous accretion rates in the thermal and kinetic modes as well as the mass accretion rate through SMBH merging as a function of redshift and for SMBHs with different final masses in Figure~\ref{fig:MassGrowthRate}. Note that, here, we do not use the subhalo merger tree, but rather the merger tree of the SMBHs themselves. For every binary BH merger, we define the more massive progenitor as the main branch, while the less massive progenitor is considered to be a contribution to the mass growth via merging. We use this tree because it is, unlike the galaxy merger tree, unambiguous and does not require additional definitions apart from the one just stated (all SMBH mergers and masses at the time of merging of the two SMBHs involved are part of the simulation output). Additionally, we show the average black hole mass as a function of redshift (black dashed line, right scale) in Fig.~\ref{fig:FractionMassGrowth} to emphasize the relative importance of different redshifts.

Low mass SMBHs are at all times dominated by the growth via accretion in the thermal mode, with mergers being a second, sub-dominant channel of growth. For SMBHs more massive than $10^{8.5}$~M$_\odot$, however, the growth in the thermal mode gets increasingly suppressed as more and more SMBHs switch to the kinetic mode. The accretion rate in the kinetic mode is always subdominant, however, even with respect to the growth by mergers. Thus, switching to kinetic mode implies that the mass growth starts to become entirely dominated by mergers, because the in-situ growth via gas accretion is reduced by up to two orders of magnitude. 

We therefore conclude that SMBHs more massive than $10^{8.5}$~M$_\odot$ grow most of their mass via mergers with lower mass black holes. The trend that mergers become more important for the mass growth of SMBHs is in qualitative agreement with other work \citep{Fanidakis+2011, Dubois+2014b}, however, it is significantly more pronounced in IllustrisTNG. This is likely caused by the very efficient feedback in our simulations, and not due to more frequent mergers per se, which is discussed further in Appendix~\ref{app:mergerrates}.

A second, interesting aspect is the average accretion rate in the kinetic mode. For SMBHs less massive than $10^9$~M$_\odot$, it reaches at most $\dot{M} \approx 10^{-3}$~M$_\odot$~yr$^{-1}$. As the accretion rate translates to a bolometric luminosity of 
\begin{align}
L_{\rm bol} \approx 1.4 \times 10^{43} {\rm erg\, s}^{-1} \frac{\dot {M}}{10^{-3}\,{\rm M}_\odot\,{\rm yr}^{-1}},
\end{align}
assuming a radiative efficiency of $0.2$ (which is highly optimistic for a SMBH accreting at low Eddington ratios), and keeping in mind that the number density of SMBHs drops steeply for SMBHs more massive than $10^{8.5}$~M$_\odot$ (not shown here), it becomes clear that AGN in the kinetic mode are not expected to play a significant role in the quasar luminosity function, i.e.~the quasar luminosity function only probes the growth in the thermal mode. 

\subsection{The quasar luminosity function}

\begin{figure*}
  \includegraphics{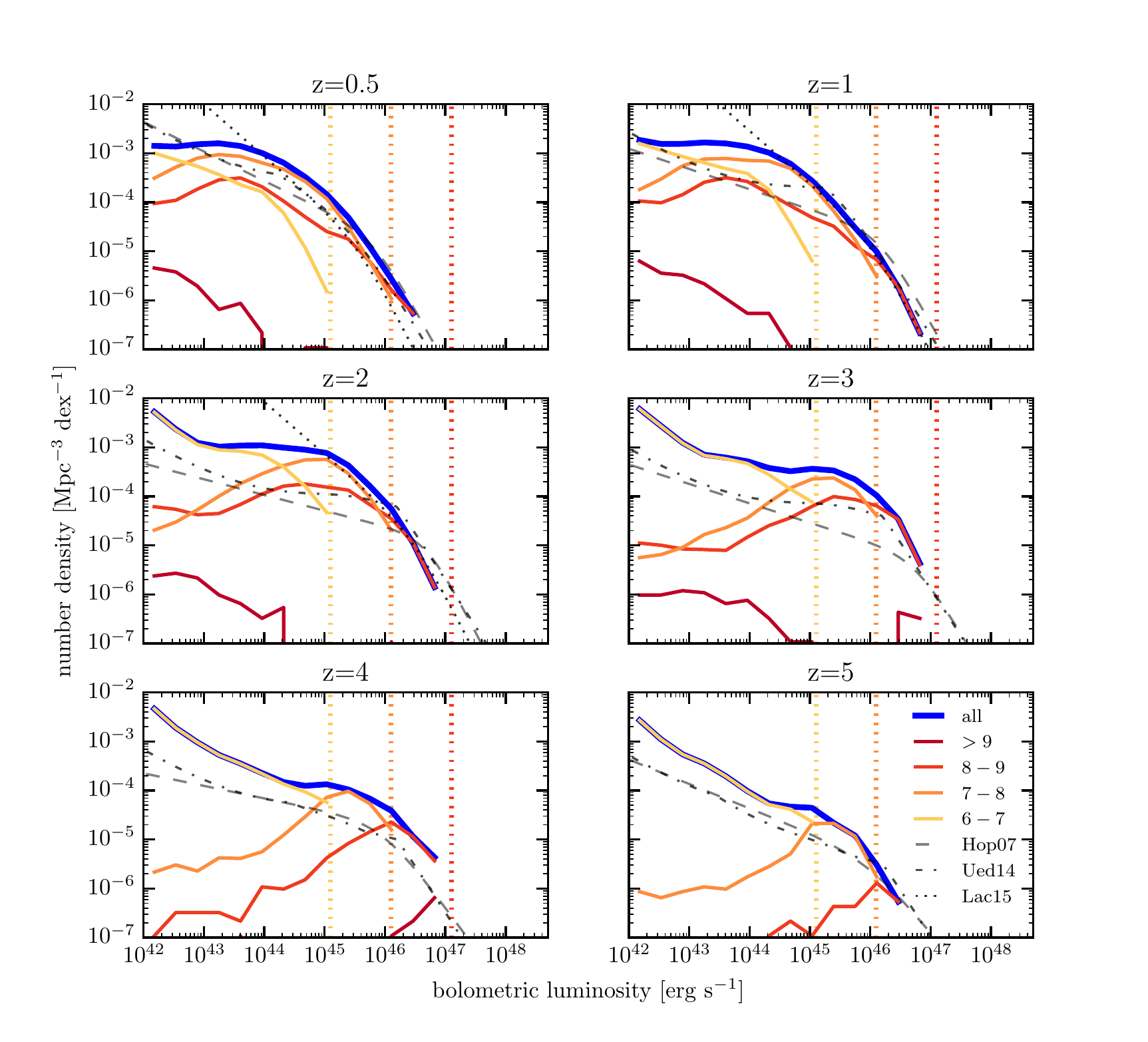}
  \caption{Quasar luminosity function at different redshifts (blue) as well as the contribution from SMBHs in different mass bins. The numbers in the legend signify the log$(M_{\rm SMBH} {\rm M}_\odot^{-1})$ limits of the respective mass bin. The dotted vertical lines show the Eddington luminosity of black holes in the respective mass bin. We also show observational fits from \citet[][Hop07]{Hopkins+2007}, \citet[][Ued14]{Ueda+2014} and \citet [][Lac15]{Lacy+2015}.}
  \label{fig:BHLF}
\end{figure*}

\begin{figure*}
  \includegraphics{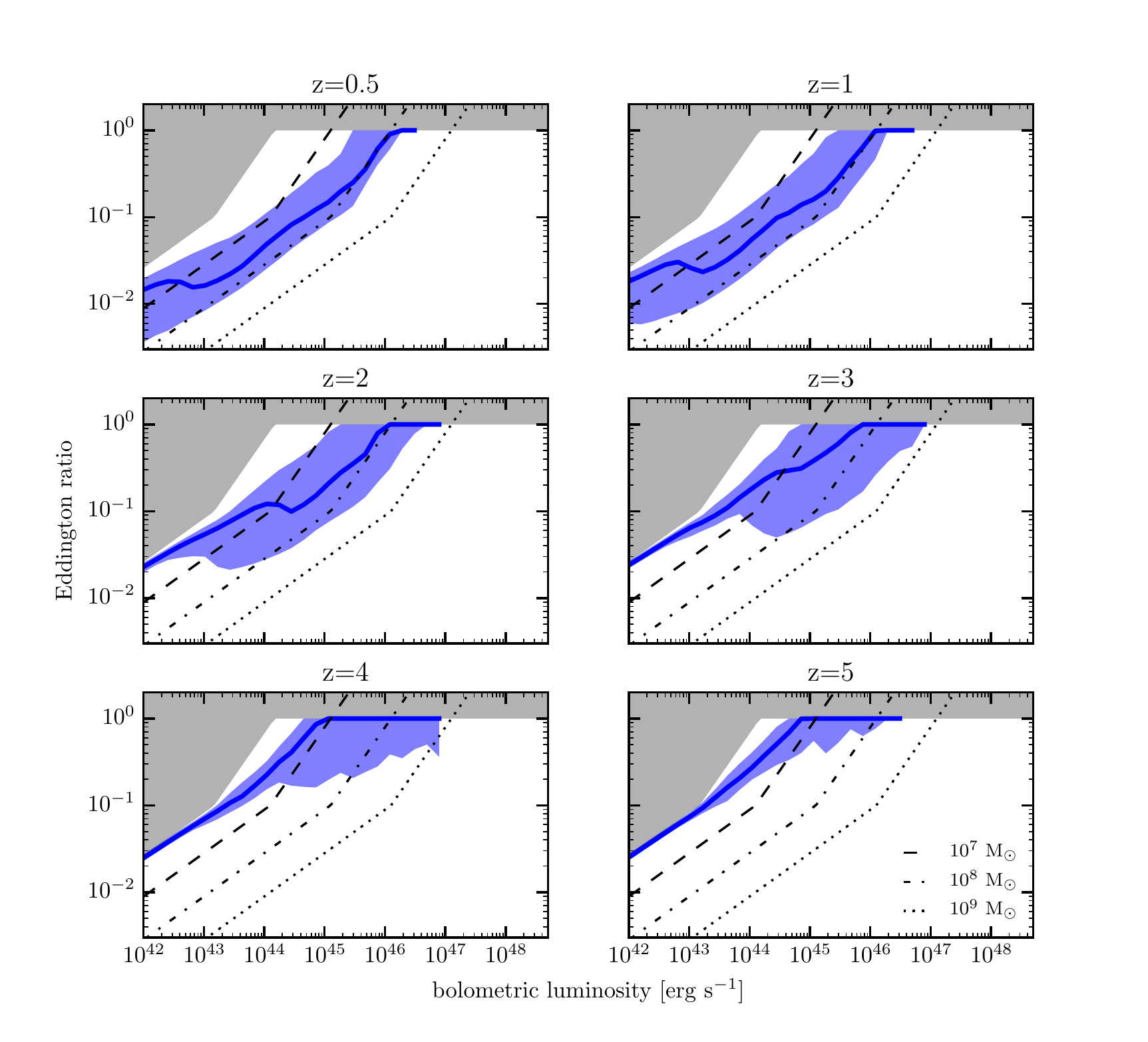}
  \caption{Median (solid line) and 10th to 90th percentile of the Eddington ratio distribution as a function of bolometric luminosity. The upper left corner (grey region) cannot be populated because our seed mass is $\sim 10^{6}$~M$_\odot$. The dashed, dash-dotted and dotted lines are lines of equal SMBH mass, $10^7$,$10^8$ and $10^9$~\Msun, respectively.}
   \label{fig:ER_LF}
\end{figure*}

The bolometric quasar luminosity function (QLF) encodes information about the instantaneous state of accretion of the SMBH population, however, in practice it just probes the most luminous black holes. In Figure~\ref{fig:BHLF}, we show the QLF at different redshifts (blue line). We also show the contribution from SMBHs  in different mass bins to facilitate the theoretical interpretation. We calculate the bolometric luminosity as
\begin{align}
L = 
\begin{cases} 
\epsilon_r\, \dot{M} c^2 \quad &\mbox{for} \quad \dot{M} \ge 0.1\, \dot{M}_{\rm Edd} , \\
 \left(10\, \frac{\dot{M}}{ \dot{M}_{\rm Edd}}\right)^2\, 0.1\, L_{\rm Edd} \quad &\mbox{for} \quad \dot{M} < 0.1\, \dot{M}_{\rm Edd},
 \end{cases}
\end{align}
assuming a decreasing radiative efficiency at low accretion rates relative to the Eddington limit \citep{Churazov+2005, Hirschmann+2014}. At high accretion rates, $\epsilon_r = 0.2$, consistent with the parameters used in the simulation. We do not model effects of obscuration, but present the QLF as a theoretical prediction of the simulation.

At low redshifts, up to $z=2$, the QLF is in good agreement with the observational fit from \citet{Lacy+2015}, overshooting at around $10^{44}$~erg s$^{-1}$ relative to \citet{Hopkins+2007, Ueda+2014}. We note however that the exact number of Compton-thick AGN at low luminosities is uncertain \citep{Buchner+2015}, thus the observational uncertainties in this regime are substantial. An additional theoretical caveat for the low-luminosity end of the QLF is that it depends significantly on the assumed radiative efficiency at low Eddington ratios \citep{Hirschmann+2014}, which might be a more complicated relation than the one assumed here \citep{Sadowski+Gaspari2017}. Note that the conversion to bolometric luminosity used here, in particular the cut at $0.1$ times the Eddington accretion rate is different from the cut used to separate the thermal from the kinetic feedback mode. It assumes that the radiative efficiency of SMBHs in the thermal mode that are accreting with lower rates than $0.1$ times the Eddington limit is lower than the value we use in the simulation. We note, however, that even assuming a constant radiative efficiency of $0.2$ for all SMBHs, which is clearly an upper limit, only affects the low luminosity end of the QLF but does not change the high luminosity regime. Keeping all the uncertainties in mind, we conclude that the simulation is in good agreement with observations at low redshift.

At high redshift, $z\geq 3$, the simulation over-predicts the QLF with respect to observations \citep{Hopkins+2007, Ueda+2014} also at the high luminosity end. We note in particular that we compare to the results from the TNG300 simulation here, i.e.~the largest volume IllustrisTNG model, because it is the only volume that probes the very rare, high-luminosity AGN. The resolution convergence of the high redshift QLF is relatively poor, with higher resolution simulations generally yielding a higher number density at fixed bolometric luminosity due to their better tracking of faster accreting black holes. This means that the discrepancy between simulation and observation at high redshift is likely alleviated by numerical resolution effects. We discuss the resolution dependence of the SMBH growth in more detail in Appendix~\ref{app:res}. 

All in all, we find that the number of luminous quasars is significantly higher in IllustrisTNG than in the Illustris simulation \citep{Sijacki+2015}, in particular at bolometric luminosities between $10^{44}$ and $10^{46}$ erg s$^{-1}$. Future observations will need to answer the question whether IllustrisTNG over-predicts the number of quasars in this luminosity range also at low redshifts, or whether the low-luminosity end of the real QLF is indeed steeper than originally inferred. Obtaining an answer to these questions will give significant constraints on SMBH seed formation and the growth of SMBHs in the low-mass regime.

To investigate the origin of the discrepancy at the high luminosity end of the high redshift QLF, we plot the contribution of SMBHs with different masses to the QLF, as well as the corresponding Eddington luminosities (vertical lines). As already indicated in the previous section, black holes more massive than $10^9$~M$_\odot$ are unimportant for the QLF. The large number of high-luminosity SMBHs at $z=3$ and $z=4$ are caused by SMBHs between $10^7$~M$_\odot$ and $10^9$~M$_\odot$, accreting at a significant fraction of the Eddington accretion rate. We confirm this with Figure~\ref{fig:ER_LF}, in which we show the median (solid blue) and $10$th and $90$th percentiles (shaded blue) of the Eddington ratio distribution vs bolometric luminosity. The grey regions show the area of the parameter space that is not allowed by our model. More than half of the SMBHs at $z=4$ that are more luminous than $10^{45}$~erg~s$^{-1}$ are experiencing Eddington-limited accretion. The only parameter that enters the Eddington luminosity is the mass of the SMBH, in particular, it is independent of the radiative efficiency. Therefore, a large number of luminous SMBHs at high redshift indicates a large number of massive SMBHs at high redshift, and an overall too early build-up of massive SMBHs at high redshift.

Comparing the Eddington ratio distribution in Figure~\ref{fig:ER_LF} with the lines of equal SMBH mass (black dashed, dash-dotted and dotted lines) shows which black holes impact different regimes of the QLF. While the high luminosity end is always dominated by black holes with masses of about $10^{8.5}$~M$_\odot$, the low luminosity end is dominated by SMBHs close to the seed mass at high redshift, and by SMBHs between $10^7$~M$_\odot$ and $10^8$~M$_\odot$ at redshift zero. This trend shows that the choice of comparably massive SMBH seeds within the model causes accretion at considerable rates after seeding in particular at high redshift, which is different compared to the scenario described in \citet{Bower+2017}.

\section{Discussion}
\label{sec:discussion}

It is a widely held conjecture that the inefficiency of star formation in high mass galaxies is caused by feedback from AGN. We show in Figure~\ref{fig:FeedbackEnergy_vs_redshift} that this is indeed the case in the IllustrisTNG simulations, where kinetic AGN feedback provides a sufficient amount of energy to balance the cooling losses of the surrounding gas and thereby maintains low star formation rates.

However, this does not answer the question what triggers this low star formation rate in the first place and steers galaxies off the star forming main sequence, as shown in Figure~\ref{fig:SFR_Mstar}. There have been a number of simulations showing that gas rich major mergers are able to drive gas to the galactic center \citep{Hernquist1989} and trigger both, a starburst and AGN quasar activity that subsequently quenches the galaxy \citep[see e.g.][]{Springel+2005,DiMatteo+2005, Hopkins+2008b, Debuhr+2012, Pontzen+2017}. However, there is no definite agreement whether this scenario is responsible for the majority of quenching events \citep[e.g.][]{Wurster+Thacker2013, Roos+2015, Sparre+Springel2017}. In IllustrisTNG, the key factor for quenching is the mass of the SMBH (Figure \ref{fig:Mbh_Mstar}) and the associated energy from kinetic AGN feedback (Figure~\ref{fig:EgyAtQuenching}). This leaves us with a large population ($>60\%$) of galaxies that are quenched unrelated to a major merger (Figure~\ref{fig:delayTime}). We note that there is an increase in quenching events that have a preceding major merger, however, it is not clear whether this is due to an increased SMBH mass growth during this merger, or whether there is a sub-population that quenches via quasar activity and switches to the kinetic mode as a consequence. In this sense, our findings do not contradict the studies of isolated systems that see this happening, but in IllustrisTNG this quenching path appears subdominant, which is also in agreement with recent observational findings \citep{Weigel+2017b}.

The SMBH model in IllustrisTNG reproduces the black hole mass--stellar mass relation (Figure~\ref{fig:Mbh_Mstar}) and does so for the following reason: low-mass SMBHs grow mainly via thermal mode accretion (Figure~\ref{fig:MassGrowthRate}), with their growth being stopped on the power-law relation due to their own feedback.  The kinetic feedback mode 
becomes important for more massive SMBH, at masses of around $10^8$~M$_\odot$ and beyond. It not only shuts off star formation, but also lowers the accretion rate by several orders of magnitude \citep[Figure~\ref{fig:MassGrowthRate} and][their Fig. 6]{Weinberger+2017}. Indeed, it lowers the accretion rate by a large enough amount that mergers of SMBHs become the dominant growth channel from this mass onwards. It has been shown \citep{Peng2007, Jahnke+Maccio2011} that hierarchical merging of galaxies and their SMBHs naturally causes a correlation close to the observed black hole mass--stellar mass relation, so this ``dry merging'' of massive black holes tends to maintain the (already established) relation. We note that this behaviour is in agreement with the scenario proposed by \citet{Graham+Scott2013}. However, the scatter in the black hole mass--stellar mass relation is smaller than in the observations and cannot be explained by the measurement errors alone.

The low accretion rates in the kinetic mode imply that SMBHs in this mode do not affect the bolometric quasar luminosity function (Figure~\ref{fig:BHLF}). Instead, the QLF probes SMBHs radiating at their Eddington luminosity at high redshift and SMBHs between $10^7$ and $10^{8.5}$~M$_\odot$ at lower redshift (Figure~\ref{fig:ER_LF}). Generally, we reproduce the low redshift QLF, which implies that the transition to a kinetic mode (for which black holes `vanish' from the QLF) is allowed by the observations of the QLF. At the high redshift regime, we over-predict the number of luminous SMBHs.
One might argue that the employed radiative efficiency of $\epsilon_r = 0.2$ is rather large, and a smaller value might lower the instantaneous luminosity. However, lowering this value self-consistently in the simulation would also lead to a more rapid mass growth \citep[][their Figure~13]{Weinberger+2017}, and consequently to an even stronger discrepancy at slightly lower redshift (as more massive black holes in general accrete at larger rates, given the same external conditions). 

A more likely explanation lies in an over-efficient early growth of the SMBHs, which might happen for a variety of reasons. One possibility is that the seeding of SMBHs happens too early and/or with too massive seeds, thereby boosting the early growth of SMBHs. An alternative solution is that some other mechanism delays the growth of low-mass SMBHs. Figure~\ref{fig:FeedbackEnergy_vs_redshift} shows that at early times, stellar feedback is energetically dominant. It has been shown in a number of studies \citep{Dubois+2014, Volonteri+2016,Angles-Alcazar+2017b,Habouzit+2017} that stellar feedback can have a large impact on the accretion rates of SMBHs. Thus, more efficient, local coupling of the stellar feedback energy at early times could delay the SMBH growth by the necessary amount and thereby lower the predicted QLF to the observed one. This possibility is in particular interesting because \citet{Grand+2017} (using a similar, but not identical model) calls for an increased early stellar feedback for a completely different reason, namely to reduce bulge-formation at high redshift in Milky Way-sized galaxies. 

The SMBH model in IllustrisTNG has been changed substantially compared to the one used in Illustris. In particular, SMBHs are seeded at higher masses, the `boost factor' in the accretion formula is abandoned, and, via a BH mass dependent threshold, it becomes more difficult for low-mass SMBHs to enter the low accretion state. All these changes affect the growth of SMBHs substantially, thus it is not surprising that the resulting QLF is different. In fact, SMBHs contribute to it very differently in IllustrisTNG compared to Illustris \citep{Sijacki+2015}. Overall, we consider the results of our comparison of the IllustrisTNG predictions for SMBHs with observations of AGN encouraging. The discrepancies we found at high redshift can help to refine the model once more detailed comparisons with observational data have been carried out, for example of the observed distributions of Eddington ratios, the occupation fractions and of the quasar clustering.

\section{Conclusion}
\label{sec:Conclusion}

The IllustrisTNG simulations reproduce a wide range of observations \citep{Genel+2017, Marinacci+2017, Naiman+2017, Nelson+2017, Pillepich+2017b, Springel+2017, Vogelsberger+2017} and arguably represent the currently best model for galaxy formation physics in hydrodynamical simulations of cosmological volumes.  We here studied the relative importance of stellar and AGN feedback channels in IllustrisTNG as a function of galaxy mass and redshift. At high redshifts, stellar feedback dominates the energy release in all galaxies. At lower redshifts, depending on galaxy mass, thermal AGN feedback takes over and becomes dominant. While the energy injected via thermal AGN feedback is formally remarkably high, only a small fraction of the energy actually acts onto the host galaxies, whereas large parts are immediately lost due to the large cooling rates of the gas surrounding the SMBHs, reducing the efficiency of this feedback channel. For massive galaxies with a redshift $z=0$ stellar mass larger than $10^{10.5}$~M$_\odot$, kinetic AGN feedback takes over at late times, which is coincident with quenching of the host galaxies and keeping them in a state of inefficient star formation.

Alongside with quenching, the kinetic feedback also self-regulates and equilibrates at accretion rates that are orders of magnitude below the average mass growth rate of mergers. This means that high mass SMBHs are formed predominantly via mergers of lower mass SMBHs. This leads to a buildup of the black hole mass -- stellar mass relation according to the scenario outlined in \citet{Graham+Scott2013}. 

Another consequence of this behaviour is that the bolometric quasar luminosity function only probes black holes in the thermal mode, i.e.~SMBHs less massive than $10^{8.5}$~M$_\odot$. In general, SMBHs in IllustrisTNG seem to grow too fast at high redshift which might have a number of reasons. One mechanism that could alleviate this discrepancy would be a more efficient (or possibly additional e.g. \citealt{Stinson+2013, Hopkins+2017}) stellar feedback channel at high redshift, as proposed by \citet{Grand+2017} for an entirely independent reason. Another would be to consider smaller SMBH seed masses, which would slow their high redshift growth substantially due to the sensitive dependence of the Bondi growth time on the black hole mass.

Finally, we note that another important consequence of our two mode model of thermal and kinetic feedback, as described in \citet{Weinberger+2017}, is a partial decoupling of both the AGN luminosity and events that might trigger AGN activity (such as mergers) from the quenching of massive central galaxies. This leads to a scenario that may make it observationally very difficult to establish a simple AGN--galaxy quenching connection. We showed in this paper how such a model behaves in a cosmological framework. A collection of other works has compared different aspects of the IllustrisTNG simulations with observational data, showing that it -- as far as we tested so far -- represents a viable scenario for galaxy formation. One critical ingredient of this model is a mass dependent switch in feedback mode at SMBH masses of around $10^8$~M$_\odot$, which is so-far only poorly motivated. Investigating possible physical explanations why such a change of modes exists will therefore be a particularly interesting topic of future research.

\section*{Acknowledgements}

The authors would like to thank Vicente Rodriguez-Gomez for generating merger trees for the simulation and Melanie Habouzit for the useful comments.  RW, VS and RP acknowledge support through the European Research Council under ERCStG grant EXAGAL-308037, and would like to thank the Klaus Tschira Foundation. RW acknowledges support by the IMPRS for Astronomy and Cosmic Physics at the University of Heidelberg. VS also acknowledges support through subproject EXAMAG of the Priority Programme 1648 `Software for Exascale Computing' of the German Science Foundation.  MV acknowledges support through an MIT RSC award, the support of the Alfred P. Sloan Foundation, and support by NASA ATP grant NNX17AG29G.
JPN acknowledges support of NSF AARF award AST-1402480.
The Flatiron Institute is supported by the Simons Foundation. 
SG and PT acknowledge support from NASA through Hubble Fellowship grants HST-HF2-51341.001-A and HST-HF2-51384.001-A, respectively, awarded by the STScI, which is operated by the Association of Universities for Research in Astronomy, Inc., for NASA, under contract NAS5-26555.
The flagship simulations of the IllustrisTNG project used in this work have been run on the HazelHen Cray XC40-system at the High Performance Computing Center Stuttgart as part of project GCS-ILLU of the Gauss Centre for Supercomputing (GCS). 
Ancillary and test runs of the project were also run on the Stampede supercomputer at TACC/XSEDE (allocation AST140063), at the Hydra and Draco supercomputers at the Max Planck Computing and Data Facility, and on the MIT/Harvard computing facilities supported by FAS and MIT MKI.






\appendix

\section{SMBH merger rates}
\label{app:mergerrates}

\begin{figure}
  \includegraphics{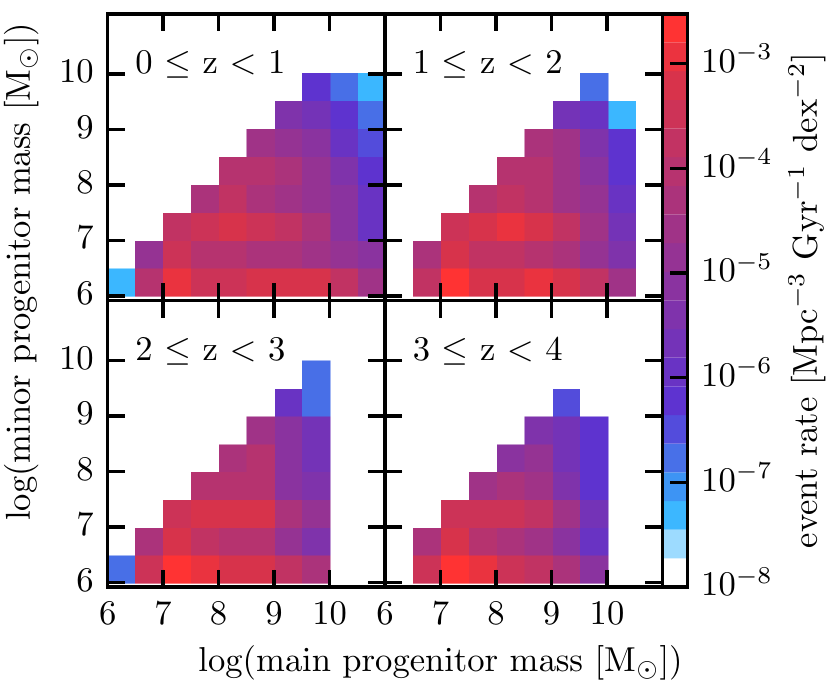}
  \includegraphics{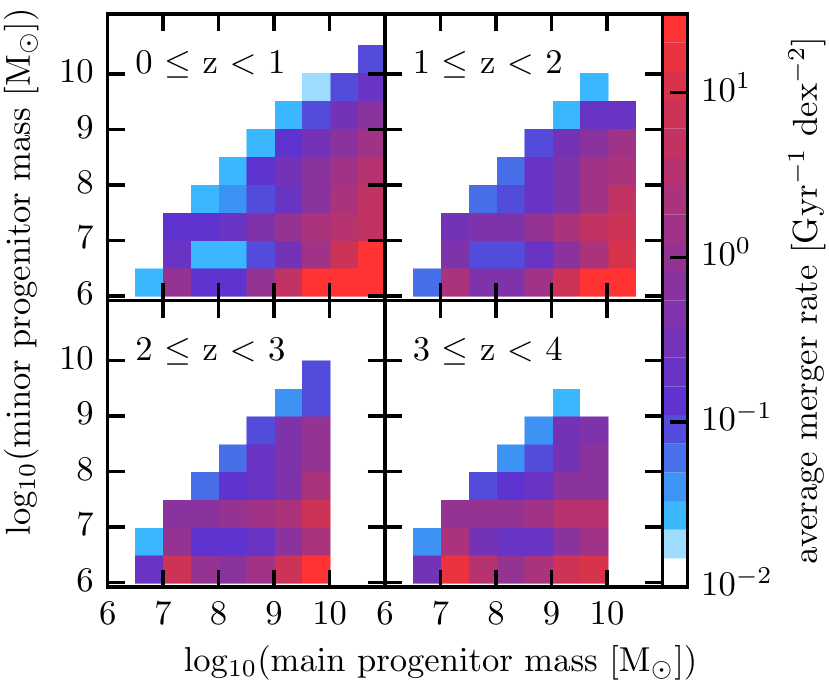}
  \caption{SMBH merger rates at different redshifts, event rate (top panel) and the average rate per SMBH (bottom panel). The rate per SMBH is calculated using the merger rate divided  by the arithmetic mean of the mass functions at the start and end of the redshift interval. Overall, the merger rates are only mildly evolving, with slightly more frequent low-mass mergers at high redshift, but fewer high-mass mergers. Even though the event rate of high mass SMBH mergers is low, the rate per SMBH is very high.}
  \label{fig:Mergers}
\end{figure}

As shown in Figures~\ref{fig:FractionMassGrowth} and \ref{fig:MassGrowthRate}, mergers of SMBHs are important for their mass growth at the high mass end. We quantify the SMBH merger rates for different massive progenitors and at different redshifts in Figure~\ref{fig:Mergers}. The top panel shows the SMBH merger event rates, which is also a prediction for gravitational wave events. We note that the simulation uses an instantaneous merger approach, i.e.~time delays between a galaxy merger and the merger of the associated SMBH pair are not modelled here and are assumed to be negligibly short. To predict the inspiral time accurately, higher resolution simulations with explicit dynamical friction treatments \citep{Tremmel+2017b} or more sophisticated post-processing analysis \citep{Kelley+2017, Kelley+2017b} would have to be employed. However, due to the remarkably constant event rate in different redshift bins, this is likely not going to alter the global event rate in a significant way.

In total, there are around $740000$ SMBH merger events in the TNG300 simulation, which corresponds to a merger density of $2.7\times 10^{-2}$~Mpc$^{-3}$ and an average event rate of $1.9\times 10^{-3}$~Mpc$^{-3}$~Gyr$^{-1}$. We note that this is about a factor of two smaller than in the $100$~Mpc EAGLE simulation, which contains $\sim 55000$ SMBH mergers \citep{Salcido+2016}. This difference can be explained by the seeding of less massive halos with lower mass SMBH in the EAGLE simulation \citep{Schaye+2015} and the resulting larger number of SMBHs.

Comparing the merger rates in Figure~\ref{fig:Mergers} with \citet[][their Figure~4]{Salcido+2016}, it is obvious that IllustrisTNG, unlike EAGLE, does not predict an excess of mergers of seed mass black holes with each other. In IllustrisTNG the merger growth is more evenly distributed, indicating that SMBH seeds do not have a delayed growth compared to their host systems, unlike in EAGLE. These differences in merger rates at black hole masses up to $10^{7.5}$~\Msun imply that future space-based gravitational wave detectors such as eLisa \citep{Amaro-Seoane2012} will be able to differentiate between different models for the SMBH growth. The need for observational constraints on this quantity can also be appreciated when comparing our inferred SMBH merger rate with the prediction of the formation rate of SMBH pairs from \citet{Tremmel+2017b}, which is $1.3\times 10^{-2}$~Mpc$^{-3}$~{Gyr}$^{-1}$ and therefore almost an order of magnitude larger than our prediction. These large discrepancies are due to the significant uncertainties about the environment in which SMBHs can form, and due to the different approaches for seeding that are adopted in different models.

The lower panel of Figure~\ref{fig:Mergers} shows the merger rate divided by the SMBH mass function of the more massive progenitor (calculated as the mean of the mass functions at the start and end point of a given redshift interval). Thus, this figure shows the average SMBH merger rate for SMBHs with a given mass. We emphasize in particular that the merger rate at the high-mass end can be quite substantial, in particular due to the contribution of mergers with small mass ratios. This means that, depending on inspiral time, high-mass SMBHs might commonly occur in binaries.

\section{Resolution dependence}
\label{app:res}

\begin{figure}
  \includegraphics{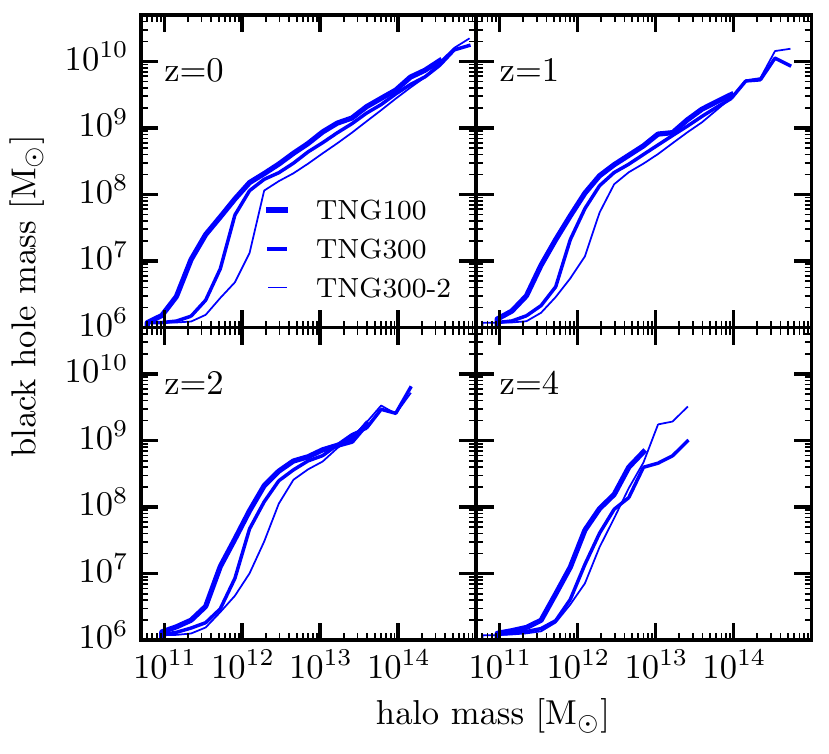}
  \includegraphics{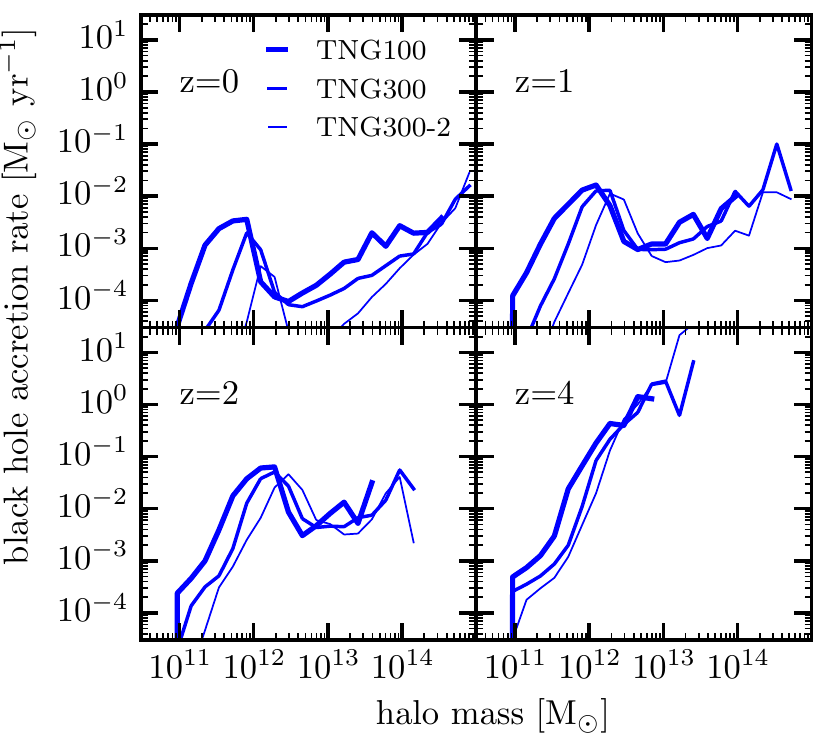}
  \caption{Resolution study with the TNG100 (high resolution), TNG300 (intermediate resolution, used in the main study) and TNG300-2 (low resolution) simulations. {\it Top panel:} Median SMBH mass as a function of host halo mass for simulations of different resolution at different redshifts. {\it Bottom panel:} Median SMBH accretion rate vs host galaxy mass. The host halo mass is a well-converged quantity, which means that the discrepancy of the simulations with different resolutions is due to the resolution dependence of the SMBH model.}
  \label{fig:Mbh_res}
\end{figure}

The predictive power of the SMBH demographics in cosmological simulations is mostly limited by the fact that they cannot model gas accretion from first principles. Consequently, sub-resolution models that estimate the accretion rate using gas properties at resolved scales need to be employed. The main problem with this approach is that the gas properties around a SMBH, i.e.~in the centre of a halo, are resolution dependent. Usually, higher densities are reached with increased resolution due to a finer sampling of the potential minimum. As a result, the estimated mass growth rate for SMBHs also increases with resolution, which leads to a more rapid growth at higher resolution. In the Bondi-Hoyle formula for the accretion rate, which is adopted in our simulations, the mass accretion rate depends on the square of the black hole mass. This means that the black hole masses of high- and low-resolution simulations will diverge with time, even in logarithmic space. This diverging behaviour can be seen not just in the time-evolution, but also in the black hole mass--total mass and the black hole accretion rate--total mass relations, which are shown in Figure~\ref{fig:Mbh_res} for different redshifts. Because a black hole is seeded at a fixed halo mass, the halo mass can be considered as a measure of the evolutionary stage of a SMBH. The masses differ significantly in the different resolution runs at intermediate halo masses, due to the effect described above.

The diverging of black hole masses is however stopped by the fact that systems will become self-regulated at a black hole mass determined by global halo and galaxy properties. This means that once this regime is reached, the black hole masses in low- and high-resolution simulations tend to converge again to a common final value. We emphasize that the black hole masses for which the resolution effect is strongest corresponds to the regime of SMBHs that contribute to the high luminosity end of the quasar luminosity function (Figure~\ref{fig:BHLF}). Therefore, we caution to over-interpret the results on the QLF, as the theoretical uncertainties are substantial.


\bsp	
\label{lastpage}
\end{document}